\documentclass[preprint,review,10pt]{elsarticle}
\usepackage{bbm}
\usepackage{amsmath,amssymb,amsthm}
\usepackage{mathrsfs}
\usepackage{color}
\usepackage{xcolor}
\usepackage{graphicx}
\usepackage{enumerate}
\usepackage{ifpdf}
\usepackage{booktabs}
\usepackage{algorithm}
\usepackage{algorithmic}
\usepackage{float}
\usepackage{CJK}
\usepackage{listings}
\usepackage{bbm}
\usepackage{mathrsfs}
\usepackage{amsmath}
\usepackage{txfonts}
\usepackage{epstopdf}
\usepackage{cases}
\usepackage{subcaption}
\usepackage{multirow}
\usepackage{enumerate}
\usepackage{enumitem}
\usepackage{lipsum}
\usepackage{xcolor}
\usepackage{float}
\usepackage{colortbl}  
\usepackage{comment}
\usepackage{pgfplots}
\usepackage{adjustbox} 
\pgfplotsset{compat=newest}
\newtheorem{assumption}{Assumption}

\usepackage{diagbox}
\makeatletter
\def\ps@pprintTitle{%
 \let\@oddhead\@empty
 \let\@evenhead\@empty
 \def\@oddfoot{}%
 \let\@evenfoot\@oddfoot}
\makeatother
 \usepackage{graphicx}
\usepackage{amssymb}
\usepackage{amsthm}
\usepackage{cancel}

\makeatletter\@addtoreset{equation}{section} \makeatother
\newtheorem{theorem}{Theorem}[section]
\newtheorem{lemma}{Lemma}[section]

\newtheorem{remark}{Remark}[section]

\begin{document}
\begin{frontmatter}
\title{Quantile regression with generalized multiquadric loss function}
\tnotetext[label1]{This work is supported by  Youth Project (Class A) of Anhui Provincial Natural Science Foundation (No. 2508085J009), NSFC (12271002).}
\author{Wenwu Gao\fnref{label1}}
 \ead{wenwugao528@163.com}
 \author{Dongyi Zheng\fnref{label1}}
 \ead{dyzheng2025@163.com}
   \author{Hanbing Zhu*\fnref{label1}}
 \ead{zhuhbecnu@163.com}
 \address[label1]{ School of Big Data and Statistics,  Anhui University, Hefei, P. R. China}
   \cortext[cor1]{Corresponding author}
\begin{abstract}
Quantile regression (QR) is now widely used to analyze the effect of covariates on the conditional distribution of a response variable.
It provides a more comprehensive picture of the relationship between a response and covariates compared with classical least squares regression.  However, the non-differentiability of the check loss function precludes 
the use of gradient-based methods to solve the optimization problem in quantile regression estimation. To this end, This paper constructs a smoothed loss function based on multiquadric (MQ) function. The proposed loss function   leads to a globally convex optimization problem that can be efficiently solved via (stochastic) gradient descent methods.  As an example, we apply the Barzilai-Borwein gradient descent method  to obtain the estimation of  quantile regression. 
We establish the theoretical results of the proposed
estimator under some regularity conditions, and compare it with other estimation
methods using Monte Carlo simulations. 
\end{abstract}
\end{frontmatter}

\thispagestyle{empty}  
\section{Introduction}\label{Intro}
\indent  Koenker and Bassett (1978) proposed the well-known quantile regression method to analyze the effect of covariates on the conditional distribution of a response variable \cite{Koenker1978RQ}. 
Let $p$ be a positive integer and $X\in \mathbb{R}^p$, $Y\in\mathbb{R}$ be two random variables with a joint distribution $F$.   Quantile regression learns the effect of $X$ on the condition distribution  of $Y$. In particular, the classical linear quantile regression reads
\begin{equation}\label{eq:linear_model}
    Q_{Y|X}(\tau) =  X^T \beta^*(\tau), \quad \tau \in (0,1),
\end{equation} where $\beta^*(\tau) \in \mathbb{R}^p$ is coefficients at quantile level $\tau$.
Furthermore, by introducing the well-known check loss function   $\rho_\tau(u) = u(\tau - \mathbb{I}(u < 0))$ with $\mathbb{I}(\cdot)$ being an indicator function, Koenker and Bassett \cite{Koenker1978RQ}  showed that   $\beta^*(\tau)$ corresponds to the minimizer of the expected risk, that is, 
\begin{equation}\label{eq:pop_obj}
    \beta^*(\tau) = \operatorname*{argmin}_{\beta(\tau) \in \mathbb{R}^p} \mathbb{E}\left[ \rho_\tau(Y - X^\top \beta(\tau)) \right]=\operatorname*{argmin}_{\beta(\tau) \in \mathbb{R}^p}\int\rho_{\tau}(t)\mathrm{d}F(t;\beta(\tau))=\operatorname*{argmin}_{\beta(\tau) \in \mathbb{R}^p}R(\beta(\tau)).
\end{equation}
Here $F(t;\beta(\tau)) := \mathbb{P}(Y - X^\top \beta(\tau) \leq t)$. In practice, they employed the empirical distribution function $F_n(t;\beta(\tau))$ based on   random samples $\{(x_i, y_i)\}_{i=1}^n$ to obtain an estimator $ \hat{\beta}(\tau)$ of $\beta^*(\tau)$ via minimizing the empirical risk, namely, 
\begin{equation}\label{eq:sample_obj_integral}
    \hat{\beta}(\tau) =\operatorname*{argmin}_{\beta(\tau) \in \mathbb{R}^p} \mathbb{E}_{F_n}\left[ \rho_\tau(Y - X^\top \beta(\tau)) \right]=\operatorname*{argmin}_{\beta(\tau) \in \mathbb{R}^p} \frac{1}{n} \sum_{i=1}^{n} \rho_\tau(y_i - x_i^\top \beta(\tau))=\operatorname*{argmin}_{\beta(\tau) \in \mathbb{R}^p}R_n(\beta(\tau)).
\end{equation}

Compared with ordinary least squares  regression, quantile regression is more robust to outliers in response measurements and heavy-tailed error distributions. Moreover, it produces  a more complete description of the conditional response distribution and uncovers different structural relationships between the response and covariates at the upper or lower tails. Early extensions focused on improving estimation robustness and flexibility, such as the nonparametric approaches by Takeuchi et al. \cite{Takeuchi2006} and the efficient composite quantile regression framework by Zou and Yuan \cite{Zou2008Composite}. In recent years, the literature has evolved to tackle more intricate data structures. Novel methodologies have been proposed to handle persistent predictors in time series (Liu et al., \cite{Liu2024}), estimate extreme conditional quantiles in nonlinear dependent processes (He and Wang, \cite{He2025108128}), and incorporate graph-structured constraints to capture spatial dependencies among predictors (Yao et al., \cite{Yao2025}). Therefore, quantile regression has been extensively studied and widely used in data science.

A  pitfall of quantile regression is that its objective function is not differentiable. Many people focus on   smoothing the objective function. Horowitz \cite{Horowitz1996BootstrapMF} smoothed the indicator component of the check function using kernel survival functions. This framework was subsequently generalized by Galvao\cite{Galvao2016SmoothQRPanel}, who relaxed the non-negativity constraint, thereby broadening the class of applicable kernel functions. Taking a different conceptual path, Fernandes et al.\cite{Fernandes2021SmoothQR} proposed smoothing the empirical distribution of the data rather than the check function itself. This alternative technique was designed to yield estimators with superior asymptotic properties, such as lower mean squared error and more accurate Bahadur-Kiefer representations. While these prominent kernel-based strategies have been instrumental in enabling differentiability, they often share a significant trade-off: the resulting smoothed objective functions are not guaranteed to be globally convex, which can complicate the search for a global minimizer.  

He et al. \cite{He2020SmoothedQR} proposed a convolution-based method to construct a twice-differentiable and convex surrogate for the quantile regression check function. These studies have significantly expanded the toolkit for QR estimation, primarily leveraging smoothing techniques to achieve differentiability, thereby enabling the application of faster gradient-based optimization algorithms. This convolution smoothing strategy has proven to be a versatile and powerful tool across various domains. In high-dimensional statistics, it was adopted by Tan et al.\cite{Tan2021Convo} to combine QR with concave regularization, effectively addressing the issues of non-smoothness and vanishing curvature to achieve oracle properties. Similarly, convolution smoothing has been successfully adapted for Support Vector Machines (SVM) by Wang \cite{WANG2024ConvoSmoothForSVM}, who transformed the non-smooth hinge loss into a differentiable surrogate. This transformation was pivotal in enabling efficient, large-scale variable selection under non-convex regularization, mirroring the computational benefits observed in smoothed quantile regression. In the context of rank regression, Zhou et al. \cite{Zhou02042024} utilized convolution smoothing to overcome the computational intractability caused by the highly non-smooth loss function in high dimensions, deriving a smooth surrogate that enables efficient and scalable estimation.  Tan et al. \cite{Tan2021Convo} utilized convolution smoothing to facilitate concave regularization, effectively transforming the piecewise linear quantile loss into a locally strongly convex surrogate that guarantees oracle properties.

However, existing convolution-based smoothing methods are often limited to providing explicit expressions only for a few specific kernels, such as the Gaussian kernel. For most other kernels, they involve numerical integration, which can be computationally intensive especially for high-dimension cases. To overcome challenges facing convolution-based smoothing methods, this paper proposes a novel technique for constructing  smooth loss functions (called GMQ function) based on multiquadric function \cite{Hardy1971}. 

Our GMQ function reads
\begin{equation}\label{GMQ}
\rho_{\tau,c}(u)=\frac{(2\tau-1)u}{2}+\frac{\sqrt{c^2+u^2}}{2}\end{equation} with $c$ being a small nonnegative shape parameter. Obviously, it includes the classical check loss function as a special case with $c=0$. More importantly, it is globally convex and infinitely smooth for any positive shape parameter $c$. This in turn leads to a    globally convex  optimization problem
\begin{equation}\label{eq:smooth_pop_obj}
    \beta_c(\tau) = \operatorname*{argmin}_{\beta(\tau) \in \mathbb{R}^p} \mathbb{E}\left[ \rho_{\tau,c}(Y - X^\top \beta(\tau)) \right] = \operatorname*{argmin}_{\beta(\tau) \in \mathbb{R}^p} \int \rho_{\tau,c}(t) \mathrm{d}F(t;\beta(\tau))=\operatorname*{argmin}_{\beta(\tau) \in \mathbb{R}^p}R_c(\beta(\tau))
\end{equation} by replacing $\rho_{\tau}$ with $\rho_{\tau,c}$ in optimization problem \eqref{eq:pop_obj}.
Moreover,  we can get an empirical estimator $\hat{\beta}_c(\tau)$  by minimizing the empirical risk 
\begin{equation}\label{eq:smooth_sample_obj_integral}
    \hat{\beta}_c(\tau) = \operatorname*{argmin}_{\beta(\tau) \in \mathbb{R}^p} \mathbb{E}_{F_n}\left[ \rho_{\tau,c}(Y - X^\top \beta(\tau)) \right] =\operatorname*{argmin}_{\beta(\tau) \in \mathbb{R}^p} \frac{1}{n} \sum_{i=1}^{n} \rho_{\tau,c}(y_i - x_i^\top \beta(\tau))=\operatorname*{argmin}_{\beta(\tau) \in \mathbb{R}^p}R_{n,c}(\beta(\tau)).
\end{equation}
Note that the  optimization problem \eqref{eq:smooth_sample_obj_integral} is smooth and globally convex. It has a unique global minimizer that can be solved efficiently with gradient-based methods.

Our construction technique has three key features. First, it is geometrically intuitive and includes the classical check loss function as a special example with a zero shape parameter. Besides, it can be readily extended to smooth other non-smooth loss functions. We take the expectile regression \cite{Newey1987AsymmetricLS} and the $k$th power expectile regression \cite{Lin2022TheKP} as two examples. Second, it leads to a globally convex optimization problem that has a unique global minimizer. The last but not the least one is that it  allows for fast computation of the unique global minimizer using (stochastic) gradient methods. More precisely,  since the gradient of the objective function only involves simple algebraic operations, it is faster to run in each iteration.  In addition, the   algebraic decay of the second derivative of our loss function yields a more robust and global estimate of the optimization problem's curvature, leading to a demonstrably more efficient convergence trajectory. 

To derive upper bounds of $|\hat{\beta}_{c}(\tau)-\beta^*(\tau)|$, we split it into two distinct parts: the smoothing bias $|\beta_c(\tau)-\beta^*(\tau)|$ and the empirical error $|\hat{\beta}_c(\tau)-\beta_c(\tau)|$. The smoothing bias arises from approximating the check loss function with GMQ function, while the empirical error  captures sampling variation.  We go further with deriving estimates of the smoothing bias as given in Lemma \ref{theoreticalerror}  and establishing the  Bahadur-Kiefer representation for the empirical error (see Theorem \ref{theorem: Bahadur}).  Both of these two theorems demonstrate that our proposed smoothing technique leads to an asymptotically unbiased coefficient estimator of linear quantile regression.

The paper is organized as follows.  Section $2$ provides the main results of the paper including GMQ loss function and its properties, theoretical analysis of linear quantile regression estimators with GMQ  loss function, and algorithms for implementing the linear quantile regression. Section $3$ provides simulations, while conclusions and discussions are provided in Section $4$.
\section{Main results}\label{section3}
\subsection{Generalized multiquadric function}\label{section3_1}
Hardy\cite{Hardy1971} first constructed the multiquadric function $\phi(x)=\sqrt{c^{2}+x^{2}}$ to smooth out the non-differentiable point $x=0$ of $|x|$. Here $c$ is a small nonnegative shape parameter. Beyond its smoothing capability, the multiquadric function has been proven to possess excellent approximation properties. Ma and Wu \cite{MA2009925, WuMa2011} demonstrated that multiquadric quasi-interpolation schemes can accurately approximate not only the target function but also its high-order derivatives, even when data points are irregularly distributed. Furthermore, compared to classical methods like divided differences, the multiquadric approach exhibits superior numerical stability and robustness, making it an efficient tool for processing scattered data with noise \cite{Ma2010Stability}.

\begin{figure*}[htbp]
  \centering
  
  \begin{subfigure}[t]{0.32\textwidth}
    \centering
    \includegraphics[width=\textwidth]{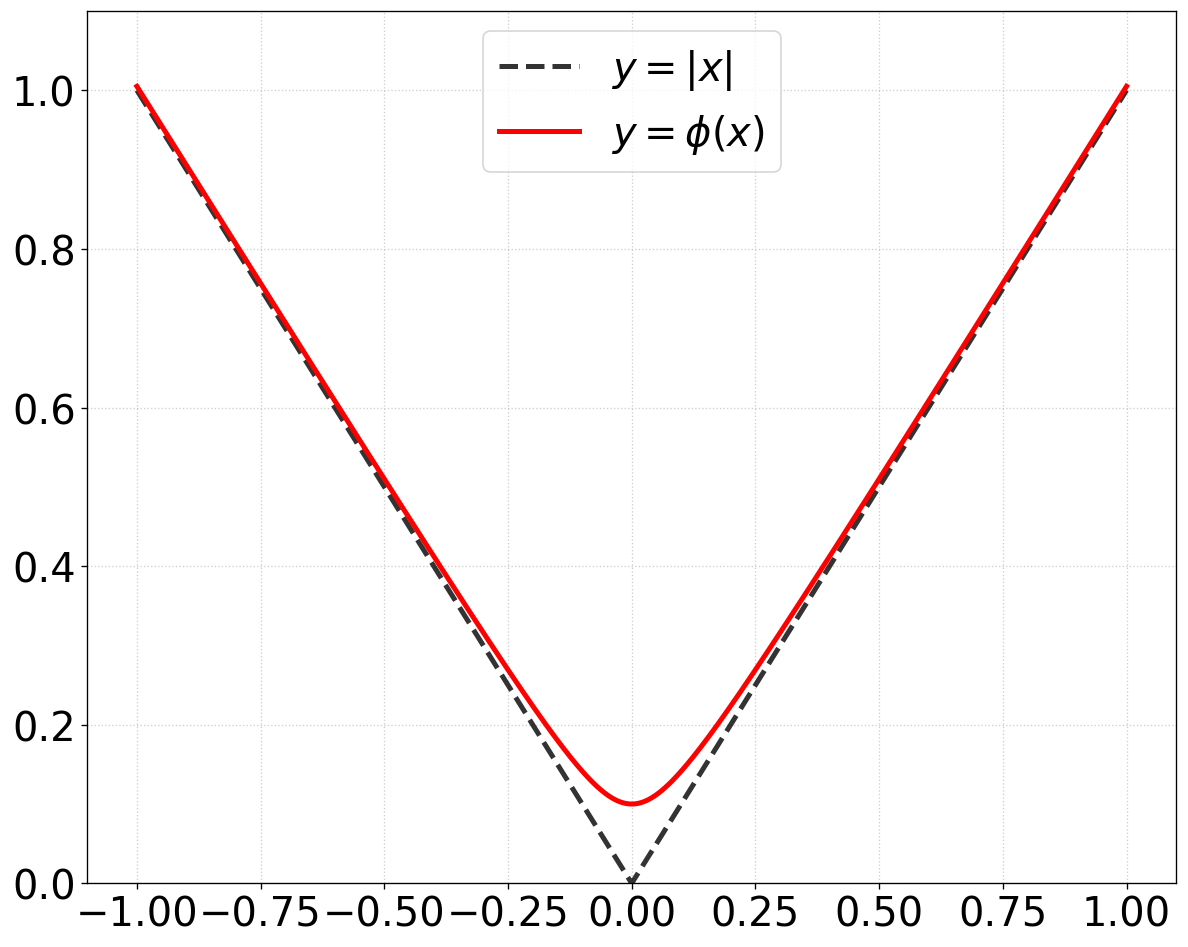}
    \caption{Absolute function and MQ function.}
    \label{fig1_1}
  \end{subfigure}%
  \hfill
  \begin{subfigure}[t]{0.32\textwidth}
    \centering
    \includegraphics[width=\textwidth]{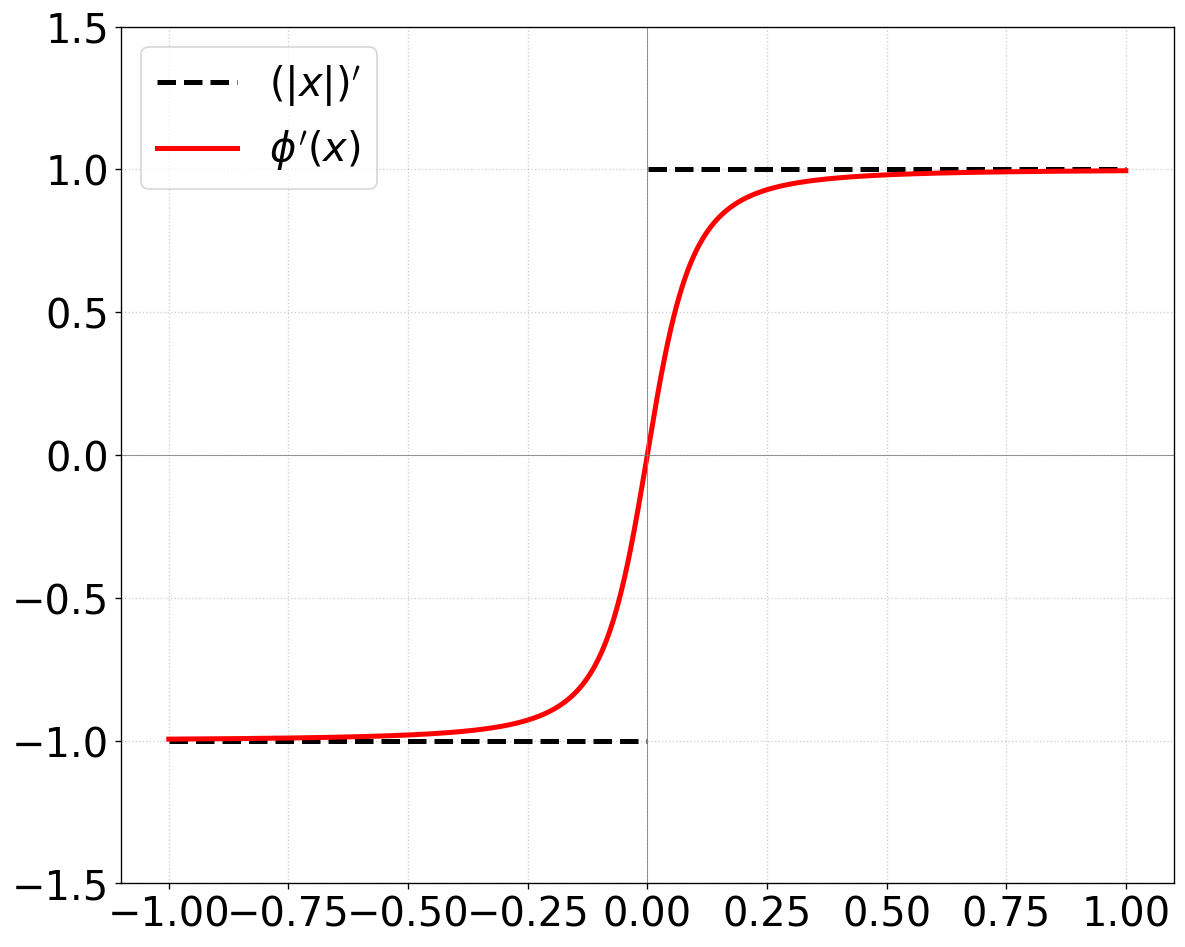}
    \caption{First-order derivative.}
    \label{fig1_2}
  \end{subfigure}%
  \hfill
  \begin{subfigure}[t]{0.32\textwidth}
    \centering
    \includegraphics[width=\textwidth]{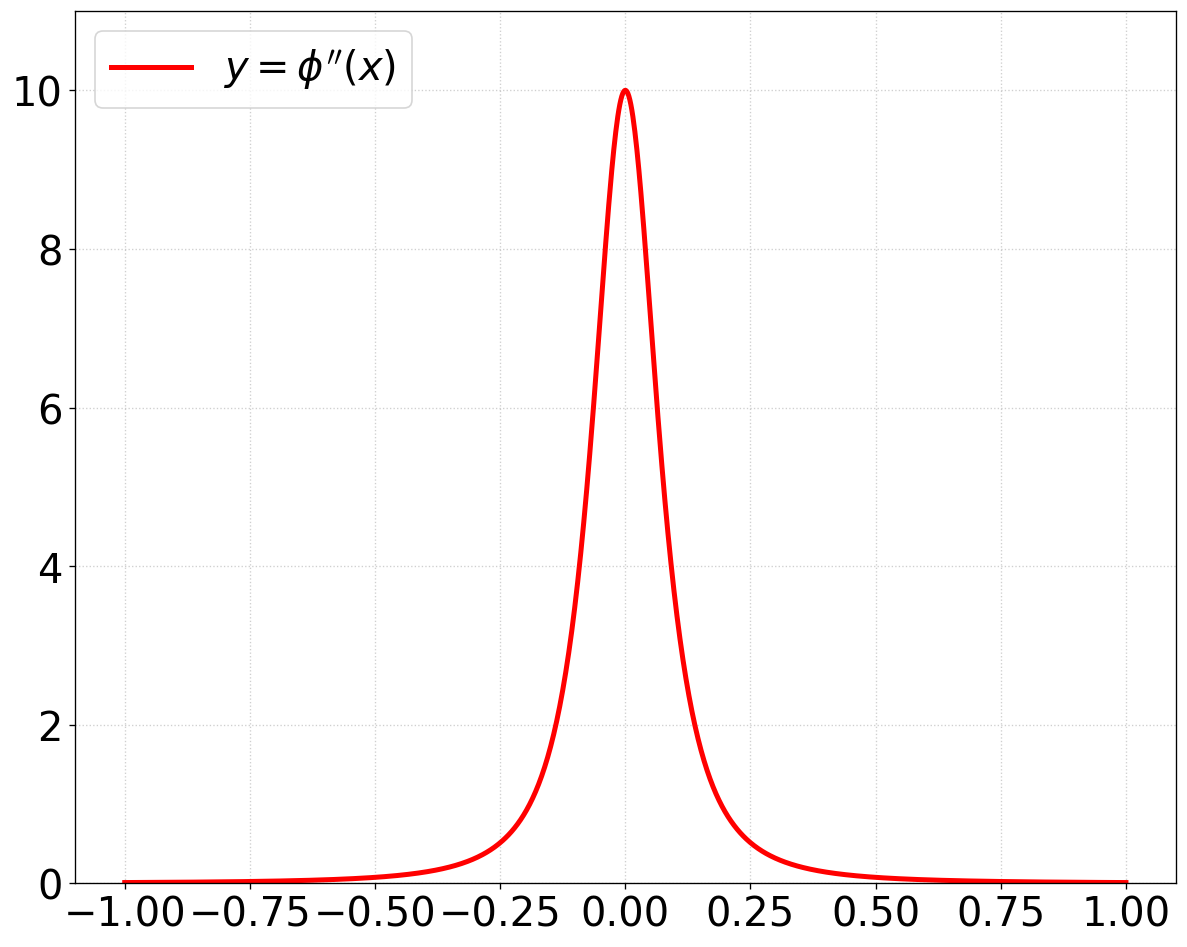}
    \caption{Second-order derivative.}
    \label{fig1_3}
    \end{subfigure}

    \caption{Absolute function, MQ function and corresponding first-order derivative, second-order derivative under $c=0.1$.}
  \label{fig1}
\end{figure*}

Here, we provide a geometric viewpoint of MQ function.  Let  $f_1(x) = x$ and $f_2(x)=-x$, then the image of $y=\phi(x)$ is the upper branch of the hyperbolas 
$$(y-f_1(x))\big(y-f_2(x)\big)=c^2.$$  Therefore, $y=\phi(x)$ approaches the two asymptote $y=f_1(x)$ and $y=f_2(x)$ as $c$ tends to zero (see Figure \ref{fig1}). Moreover importantly, $\phi(x)$ is infinitely smooth and its derivatives can be provided explicitly, for example, $$\phi^{'}(x)=\frac{x}{\sqrt{c^2+x^2}},\quad \phi^{''}(x)=\frac{c^2}{(\sqrt{c^2+x^2})^3}.$$
Such a geometric viewpoint of $\phi(x)$  will facilitate us to construct generalized multiquadric function  from the check loss function for quantile regression.

\begin{figure*}[htbp]
  \centering
  
  \begin{subfigure}[t]{0.32\textwidth}
    \centering
    \includegraphics[width=\textwidth]{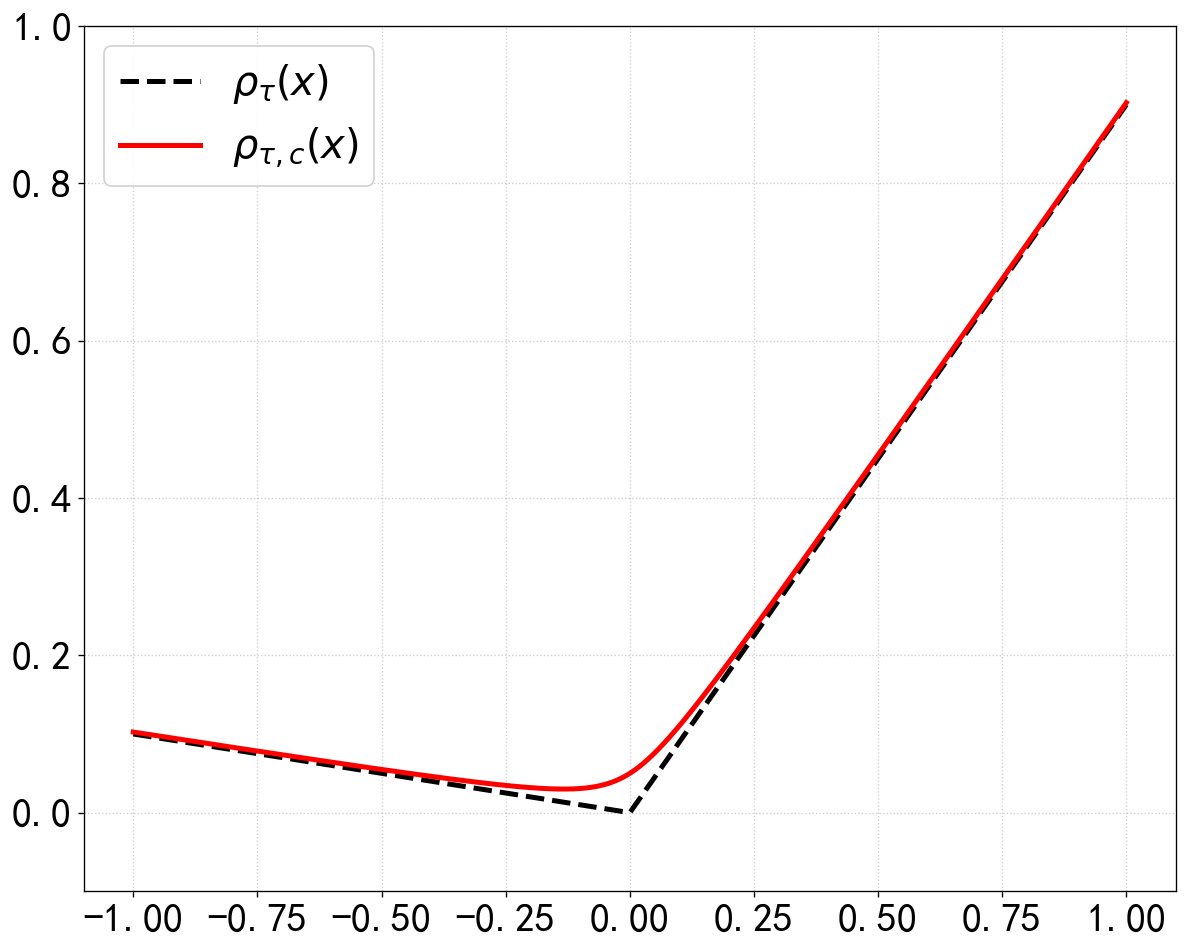}
    \caption{Check loss function and GMQ function.}
    \label{fig2_1}
  \end{subfigure}
  \hfill
  \begin{subfigure}[t]{0.32\textwidth}
    \centering
    \includegraphics[width=\textwidth]{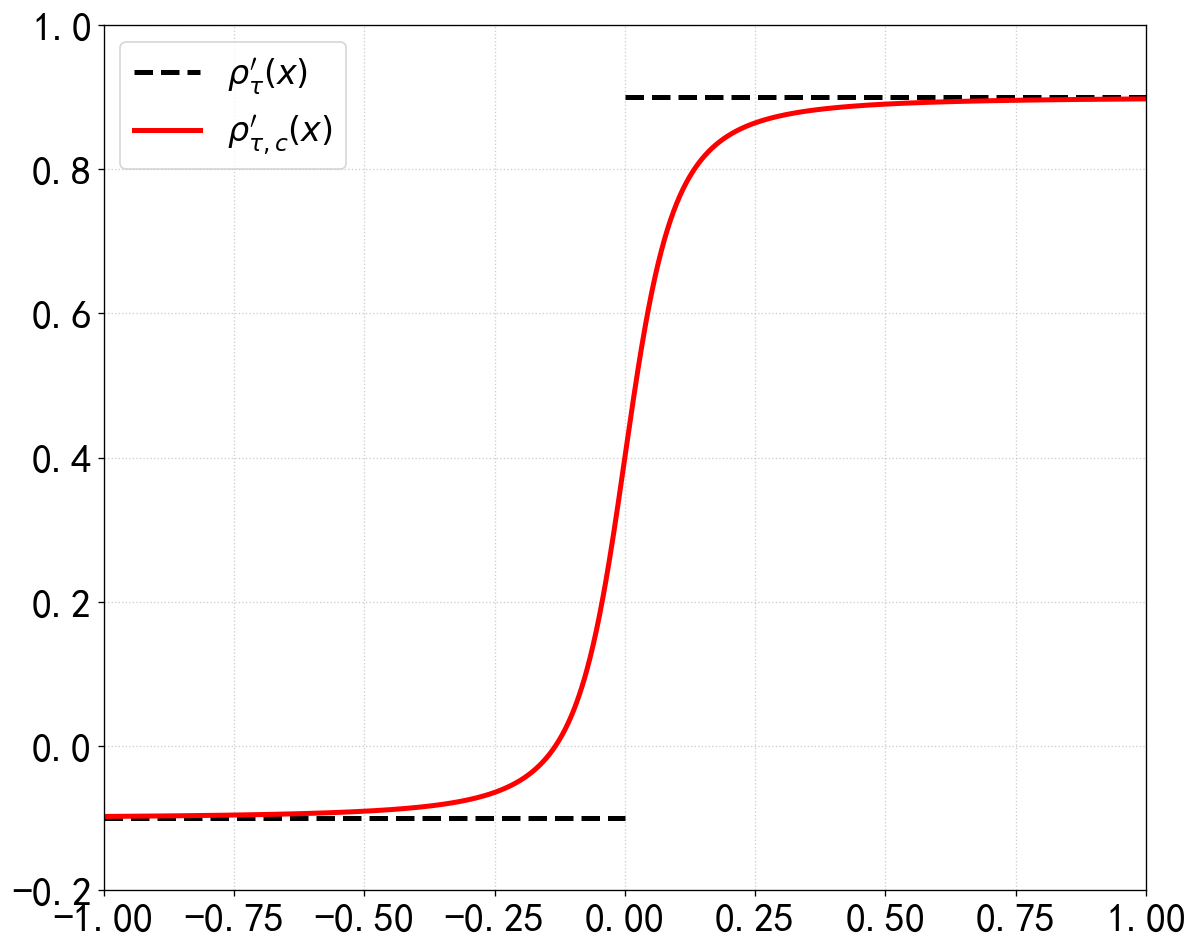}
    \caption{First-order derivative.}
    \label{fig2_2}
  \end{subfigure}
  \hfill
  \begin{subfigure}[t]{0.32\textwidth}
    \centering
    \includegraphics[width=\textwidth]{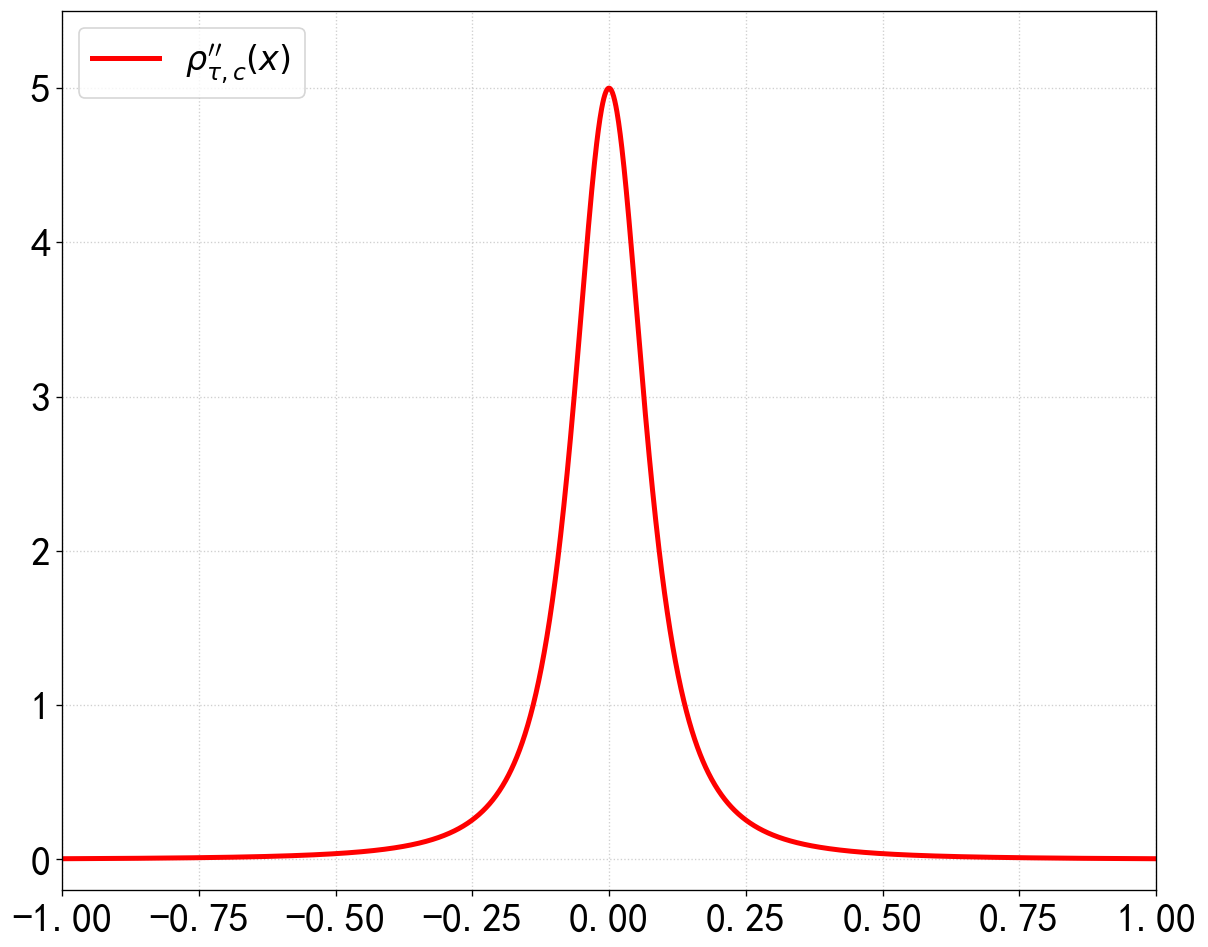}
    \caption{Second-order derivative.}
    \label{fig2_3}
  \end{subfigure}

  \caption{Check loss function, GMQ function and corresponding first-order derivative, second-order derivative under $\tau=0.9$, $c=0.1$.}
  \label{fig2}
\end{figure*}

 Let  $g_1(x) = \tau x$ and $g_2(x)=(\tau-1)x$, then the upper branch of the hyperbolas
$$(y-g_1(x))\big(y-g_2(x)\big)=c^2$$ reads $$y=\frac{(2\tau-1)x+\sqrt{c^2+x^2}}{2}=:\rho_{\tau,c}(x).$$  
This implies that the image of the GMQ function  $\rho_{\tau,c}(x)$ defined in formula \eqref{GMQ} can be viewed as a upper branch of the above hyperbolas and thus approaches its two asymptote $y=g_1(x)$ and $y=g_2(x)$ as $c$ tends to zero (see Figure \ref{fig2_1}). Therefore, it provides a smooth alternative for the check loss function $\rho_{\tau}$.

We then derive some properties of $\rho_{\tau,c}$. 
 We first explore its relation to $\phi$. Note that $\rho_{\tau}$ can be rewritten   as 
\begin{equation}\nonumber
  \rho_\tau(x)=\frac{(2\tau-1)x+|x|}{2}.
\end{equation}
This in turn leads to 
\begin{equation}\label{smoothedMQLoss}
  \rho_{\tau,c}(x)=\frac{(2\tau-1)x+\phi(x)}{2}.
\end{equation}
Consequently, we have the  identities: $$ \rho_{\tau,c}'(x)=\frac{2\tau-1}{2}+\frac{\phi^{'}(x)}{2}=\frac{2\tau-1}{2}+\frac{x}{2\sqrt{c^2+x^2}},$$ and $$ \rho_{\tau,c}^{''}(x)=\frac{\phi^{''}(x)}{2}=\frac{c^2}{2(\sqrt{c^2+x^2})^3}.$$ Moreover, with some simple derivations, it is easy to get the following lemma. \begin{lemma}\label{boundoffunctionerror} Let $\rho_{\tau}$ and $\rho_{\tau,c}$ be defined as above. Then we have   
\begin{equation}\label{errorofsmooth}
\rho_{\tau,c}(x)-\rho_{\tau}(x)\leq \begin{cases}c/2, x\leq c,\\c^2/(2|x|),   x\geq c.\end{cases}
\end{equation}
\end{lemma}
The above discussions demonstrate that  GMQ function inherit fair properties of MQ function such as smoothness, convexity, boundedness of the first-order derivative, and algebraic decaying of high-order derivatives (see Figure \ref{fig2_2}). In particular, its second-order derivative is a strictly positive definite function.  
Moreover, by replacing $\rho_\tau$ with $\rho_{\tau,c}$ in the optimization problem \eqref{eq:smooth_pop_obj}, we get an estimator $\beta_c(\tau)$ of  $\beta(\tau)$ by solving the optimization problem
\begin{equation*}
    \beta_c(\tau) = \operatorname*{argmin}_{\beta(\tau) \in \mathbb{R}^p} \mathbb{E}\left[ \rho_{\tau,c}(Y - X^\top \beta(\tau)) \right] = \operatorname*{argmin}_{\beta(\tau) \in \mathbb{R}^p} \int \rho_{\tau,c}(t) \mathrm{d}F(t;\beta(\tau)).
\end{equation*}
In practice, if we have realizations $\{(x_i,y_i)\}_{i=1}^n$ of random samples $\{(X_i,Y_i)\}_{i=1}^n$  at hand, the we can get an empirical estimator $\hat{\beta}_c(\tau)$ that is the unique global minimizer of  the empirical risk:
\begin{equation*} 
    \hat{\beta}_c(\tau) = \operatorname*{argmin}_{\beta(\tau) \in \mathbb{R}^p} \mathbb{E}_{F_n}\left[ \rho_{\tau,c}(Y - X^\top \beta(\tau)) \right] = \operatorname*{argmin}_{\beta(\tau) \in \mathbb{R}^p} \frac{1}{n} \sum_{i=1}^{n} \rho_{\tau,c}(y_i - x_i^\top \beta(\tau)).
\end{equation*}
Moreover, since the above optimization problem is globally convex with a smoothed loss function, we can employ (stochastic) gradient methods to give a fast computation of the estimator $\hat{\beta}_c(\tau)$. 
More importantly, such a geometric construction technique can be extended to construct some other smoothed loss functions. As examples, we  consider smoothing   loss functions of the $k$th ($1<k<2$) power expected regression \cite{Lin2022TheKP} and the asymmetric regression \cite{Newey1987AsymmetricLS} using the above geometric technique.
  
  Let the loss function of $k$th   power expectile regression be given as 
\begin{equation*}   \rho_\tau^{e}(x)=\begin{cases}\tau x^k,x\geq0,\\(1-\tau)(-x)^k,x<0,\end{cases} \tau \in (0,1).
\end{equation*} 
 Then  we can construct a smooth counterpart of $\rho_\tau^{e}$ in the form
\begin{equation} \nonumber
  \rho_{\tau,c}^{e}(x)=\frac{(2\tau-1)x^k+\sqrt{c^{2}+x^{2k}}}{2}.
\end{equation}
Moreover, we can verify that  $f_1(x)=\tau x^k$ and $f_2(x)=(1-\tau)(-x)^k$ are  two corresponding asymptotic functions. We go further with smoothing the  loss function of  asymmetric least squares regression.
Let
\begin{equation} \nonumber
  \rho_\tau^{as}(x)=\begin{cases}\tau x^2,x\geq0,\\(1-\tau)x^2,x<0,\end{cases}   \quad \tau \in (0,1).
\end{equation} It is easy to verify that $(\rho_\tau^{as})'(x)=2\rho_\tau(x)$. Therefore, by replacing  $\rho_\tau(x)$ with $\rho_{\tau,c}(x)$ and taking indefinite integral, we have
\begin{equation} \nonumber
  \rho_{\tau,c}^{as}(x)=\frac{(2\tau-1)x^{2}}{2}+\frac{x\sqrt{c^{2}+x^{2}}+c^{2}\ln|x+\sqrt{c^{2}+x^{2}}|}{2}-\frac{c^{2}\ln c}{2},
\end{equation} 
which is a smooth alternative of $\rho_\tau^{as}(x)$.

\subsection{Linear quantile regression with GMQ loss function}\label{section3_2}
Based on the above constructed GMQ loss function, this section aims at deriving a linear quantile regressor $Y=X^T\hat{\beta}_{c}(\tau)$ by solving the 
globally convex optimization problem 
\begin{equation*}
    \hat{\beta}_c(\tau) = \operatorname*{argmin}_{\beta(\tau) \in \mathbb{R}^p} \frac{1}{n} \sum_{i=1}^{n} \rho_{\tau,c}(y_i - x_i^\top \beta(\tau)).
\end{equation*} 

Before presenting the main theorems and lemmas, we introduce the following necessary assumptions.
\begin{assumption}\label{assumption:variableX}
    The components of $X$ are bounded random variables and the matrix $\mathbb{E}[XX']$ is full rank.
\end{assumption}
\begin{assumption}\label{assumption:densityFunction}
    The density function $f(\cdot)$ is bounded, strictly positive, and continuously differentiable. Furthermore, its first derivative $f'(\cdot)$ is uniformly bounded.
\end{assumption}

To derive bounds of regression error, we only need to derive the ones of $|\hat{\beta}_c(\tau)-\beta^*(\tau)|$ due to the linear structure  of the regressor. We first split $|\hat{\beta}_c(\tau)-\beta^*(\tau)|$ into two parts: $|\hat{\beta}_c(\tau)-\beta_c(\tau)|$ and $|\beta_c(\tau)-\beta^*(\tau)|$. Moreover, based on the triangle inequality, we have $$|\hat{\beta}_c(\tau)-\beta^*(\tau)|\leq |\beta_c(\tau)-\beta^*(\tau)|+|\hat{\beta}_c(\tau)-\beta_c(\tau)|.$$  
Observe that 
\begin{equation*}
    \beta^*(\tau) =\operatorname*{argmin}_{\beta(\tau) \in \mathbb{R}^p}\int\rho_{\tau}(t)\mathrm{d}F(t;\beta(\tau))
\end{equation*}
and 
\begin{equation*}
    \beta_c(\tau) =\operatorname*{argmin}_{\beta(\tau) \in \mathbb{R}^p}\int\rho_{\tau,c}(t)\mathrm{d}F(t;\beta(\tau)).
\end{equation*} In addition, since the above two optimizations problems are globally convex,  $\beta^*(\tau)$ and $\beta_c(\tau)$ are unique. Therefore, the error $|\beta_c(\tau)-\beta^*(\tau)|$ is completely characterized by
$\int|\rho_{\tau}(t)-\rho_{\tau,c}(t)|\mathrm{d}F(t;\beta(\tau))$. 
Then we can get the following lemma.

\begin{lemma}\label{theoreticalerror}
Let $\beta_c(\tau)$ and $\beta^*(\tau)$ be  defined as above. Assume that the density function $f$ is a bounded continuous function. Then we have the error estimate
\begin{equation}
|\beta_c(\tau)-\beta^*(\tau)|=\mathcal{O}(c^2|\ln c|),\  \ \tau\in (0, 1).
\end{equation}
\end{lemma}
Proof of Lemma \ref{theoreticalerror} see appendix.
We go further with deriving the bound of the error $|\hat{\beta}_c(\tau)-\beta_c(\tau)|$. 
\begin{lemma}\label{lemma:stochastic_order}
 Let $\hat{\beta}_c(\tau)$ and $\beta_c(\tau)$ be  defined as above. Then, under Assumptions \ref{assumption:variableX} and \ref{assumption:densityFunction}, we have 
\begin{equation}\nonumber
  \|\hat{\beta}_c(\tau) - \beta_c(\tau)\| = O_p\left( \frac{1}{\sqrt{n}} \right).
\end{equation}
\end{lemma}
Proof of Lemma \ref{lemma:stochastic_order} see appendix.

The above Lemma \ref{lemma:stochastic_order} establishes the $\sqrt{n}$-consistency of the smoothed estimator $\hat{\beta}_c(\tau)$ to the parameter $\beta_c(\tau)$. This together with Lemma \ref{theoreticalerror} yields following theorem.
\begin{theorem}\label{lemma:rateOfConvergenceWithProb}
Let the assumptions of Lemma \ref{theoreticalerror} and Lemma \ref{lemma:stochastic_order} hold. The smoothed quantile regression estimator $\hat{\beta}_c(\tau)$ converges in probability to the true parameter $\beta^*(\tau)$ with the rate:
\begin{equation}
    \left\|\hat{\beta}_c(\tau) - \beta^*(\tau)\right\| = O_p\left( n^{-1/2} + c^2 |\ln c| \right).
\end{equation}
\end{theorem}

 Following the rigorous framework established in \cite{Fernandes2021SmoothQR}, we only present the following three  theorems without proof, readers are referred to  the reference \cite{Fernandes2021SmoothQR}  for the comprehensive proof techniques. The next theorem derives some convenient expansions for the stochastic error $\hat{\beta_c}(\tau)-\beta_c(\tau)$.   For this purpose, let $S_{n,c}(\tau):=\nabla R_{n,c}\big(\beta_c(\tau)\big)$ and $D_c(\tau):= \nabla^2R_c\big(\beta_c(\tau)\big) $.  Note that the first order condition $\nabla R_c\big(\beta_c(\tau)\big)=\mathbf{0}$ implies that the score term $S_{n,c}(\tau)$ has zero mean, and hence, the stochastic error in the Bahadur–Kiefer representation (Theorem \ref{theorem: Bahadur}) is asymptotically centered.

\begin{theorem}\label{theorem: Bahadur}
Under Assumptions \ref{assumption:variableX} and \ref{assumption:densityFunction}, with probability approaching one, the estimator satisfies the following representation:
\begin{equation}\nonumber
  \sqrt{n}\left(\hat{\beta}_{c}(\tau)-\beta_{c}(\tau)\right) = -\sqrt{n} D_{c}^{-1}(\tau)S_{n,c}(\tau) + O_p(\varrho_{n}(c)),
\end{equation}
  where $\varrho_n(c) = \sqrt{\frac{\ln n}{nc}}$. 
\end{theorem}

\indent let $\Sigma_c(\tau) := Var(\sqrt{n} D_c^{-1}(\tau) S_{n,c}(\tau))$ denote the asymptotic covariance matrix of the smoothed estimator. The following theorem characterizes the asymptotic covariance matrix $\Sigma_c(\tau)$ of the smoothed estimator and explicitly quantifies its efficiency gain over the standard QR estimator.
\begin{theorem}\label{theorem:variance_reduction} 
  Under Assumptions \ref{assumption:variableX} and \ref{assumption:densityFunction}., the asymptotic covariance matrix of $\hat{\beta}_c(\tau)$ admits the expansion:
\begin{equation}\nonumber
    \Sigma_c(\tau) = \Sigma(\tau) - \frac{\pi}{4} c D^{-1}(\tau) + o(c),
\end{equation}
where $\Sigma(\tau) = \tau(1-\tau)D^{-1}(\tau)E[XX^T]D^{-1}(\tau)$ is the asymptotic covariance matrix of the standard QR estimator, $D(\tau) = E[XX^Tf(X^T\beta^*(\tau)|X)]$ is the Hessian matrix.
\end{theorem}
Theorem \ref{theorem:variance_reduction} provides a strong theoretical justification for smoothing. It demonstrates that the asymptotic covariance $\Sigma_c(\tau)$ is reduced relative to $\Sigma(\tau)$ by a term proportional to $c$. This implies that, beyond computational benefits, the smoothed estimator strictly dominates the standard QR estimator in terms of asymptotic efficiency for small $c$.

\indent With the expansions for the bias and the variance in hand, we can now derive the theoretically optimal value for the smoothing parameter $c$. This optimal value, denoted $c^*$, is chosen to minimize the Asymptotic Mean Squared Error ($AMSE\big(\lambda^T\hat{\beta}_c(\tau)\big)=\mathbb{E}\!\left[\lambda^T\big(\beta_c(\tau)-D_c^{-1}(\tau)S_{n,c}(\tau)-\beta^*(\tau)\big)\right]^2 = \left( \text{Bias}(\lambda^T\hat{\beta}_c) \right)^2 + \text{Var}(\lambda^T\hat{\beta}_c))$ of the estimator for a specific linear combination of the coefficients, $\lambda^T\hat{\beta}_c(\tau)$.
\begin{theorem}\label{theorem:optimal_c} 
  Let Assumptions \ref{assumption:variableX} and \ref{assumption:densityFunction} hold. If $\lambda^TB(\tau) \neq 0$, then the $AMSE(\lambda^T\hat{\beta}_c(\tau))$ is minimized for:
$$c^*_\lambda = \left( \frac{\pi/4 \cdot \lambda^TD^{-1}(\tau)\lambda}{4n [B(\tau)]^2} \right)^{1/3}$$
where $B(\tau) = \frac{1}{2} D^{-1}(\tau) \mathbb{E}\left[ X f_Y^{(1)}\big(X^T\beta(\tau) \mid X\big) \right]$, and $D(\tau)=\nabla^2R(\beta^*(\tau))$.
The resulting minimal AMSE is equal to:
$$ AMSE(\lambda^T\hat{\beta}_{c^*}(\tau)) = \frac{1}{n} \lambda^T \left[ \Sigma(\tau) - \frac{3\pi}{16}  c^*_\lambda D^{-1}(\tau) \right] \lambda + o(c^*/n) $$
\end{theorem}
\indent Theorem \ref{theorem:optimal_c} establishes the explicit expression for the asymptotically optimal smoothing parameter $c^*$. Its $n^{-1/3}$ rate of convergence is analogous to the optimal bandwidth for kernel-based methods using a second-order kernel. Although a direct "plug-in" estimation of $c^*$ is non-trivial due to the dependence of $B(\tau)$ on unknown derivatives of the density function, the theorem provides a robust theoretical foundation for data-driven selection strategies, such as cross-validation. In particular, it theoretically justifies the adoption of an $n^{-1/3}$ scaling law when constructing practical rules of thumb for $c$.

\subsection{Algorithm implementation}\label{section3_3}
\indent The theoretical necessity of smoothing arises from the superior analytical and computational properties of smooth functions compared to their nonsmooth counterparts. The continuity of first-order derivatives in smooth functions not only facilitates the use of tools like Taylor expansions but also simplifies theoretical modeling through concise expressions. Consequently, smooth functions are fundamental in machine learning and optimization. For instance, in loss function minimization, smoothness guarantees clear gradient information, allowing gradient descent to readily identify local minima—a process that is considerably more arduous with nonsmooth functions.
Thus, smoothing is instrumental to the efficacy of generalized MQ functions in regression tasks. Algorithmically, generalized MQ functions integrate two asymptotic functions using an improved double cubic Hermite interpolation. This method enforces derivative consistency at the connection points, successfully balancing smoothness with high approximation accuracy.
\\ \indent In the previous section, we constructed the smooth generalized MQ function based on the idea of asymptotic lines and hyperbolas, and thus constructed a smooth loss function (\ref{smoothedMQLoss}) for quantile regression. We need to optimize the objective function $R_{n,c}(\beta(\tau))$, among $\beta =(\beta_1, \beta_2, ..., \beta_p)$ the parameters of the model. For the sake of convenience, we will refer to $\beta(\tau)$ as $\beta$ in the following text.  Vanilla gradient descent (GD) is the most basic gradient descent algorithm that can be used to optimize the objective functions of various models. The key idea of GD is to compute the gradient of the objective function $R_{n,c}(\beta)$ with respect to the parameter $\beta$, and then update the parameter along the opposite direction of the gradient, in order to minimize the objective function. Specifically, given an initialized $\beta^0 \in \mathbb{R}^p$, shape parameter $c$, at iteration $t=0,1,2,3,...$, the GD update rule is:
\begin{equation}\nonumber
    \beta^{t+1}=\beta^t-\eta_t \cdot \nabla R_{n,c}(\beta^t)=\beta^t-\frac{\eta_t}{n}\Sigma_{i=1}^{n}\{\rho_{\tau,c}'(y_i-x_i^T\beta^t)\}
\end{equation}
where $\eta_t>0$ controls the step size of each iteration update. This algorithm iteratively computes gradients and updates parameters until the parameter $\beta$ gradually approximates a local minimum of the objective function. In classical GD, a line search technique is usually used to obtain the step size. However, for large-scale settings, line search is computationally expensive. One of the most important issues in GD is determining a proper update step size and decay schedule. Common practices in the literature are using a decaying step size or optimally tuned fixed step size. But these all have their flaws. In this paper, we use Barzilai-Borwein (BB) gradient descent with adaptive step size \cite{Barzilai1988TwoPointSS} to solve generalized MQ quantile regression, guided by the quasi-Newton method. BB has been shown to be an effective approach for solving nonlinear optimization problems. The BB method is defined as follows: 
\begin{equation}\nonumber
    \eta_{1,t}=\frac{<\delta^t,\delta^t>}{<\delta^t,g^t>}, \eta_{2,t}=\frac{<\delta^t,g^t>}{<g^t,g^t>}
\end{equation}
Where:
\begin{equation}\nonumber
    \delta^t=\beta^t-\beta^{t-1}, g^t=\nabla R_{n,c}(\beta^t)-\nabla R_{n,c}(\beta^{t-1}),t=1,2,....
\end{equation}
\indent Therefore, the iterative process of the BB algorithm is as follows: 
\begin{equation}\nonumber
    \beta^{t+1}=\beta^t-\eta_{m,t}\cdot\nabla R_{n,c}(\beta^t),m=1 ~or~ 2.
\end{equation}
\indent The BB algorithm starts from iteration 1. At the initialization, we take a random initial value $\beta^0$, then use standard gradient descent to compute the parameter $\beta^1$. See Algorithm 1 for the detailed steps. 

\begin{algorithm}[htb] 
	\caption{Gradient descent with Barzilai-Borwein step size (GD-BB) for solving generalized MQ quantile regression.}
	\label{alg1}
	
	\renewcommand{\algorithmicrequire}{\textbf{Input:}}
	\renewcommand{\algorithmicensure}{\textbf{Output:}}
	
	\begin{algorithmic}[1] 
		\REQUIRE Data points $\{(x_i, y_i)\}_{i=1}^n$, quantile level $\tau \in (0,1)$, smoothing parameter $c \in (0,1)$, initial parameter $\beta^0$, convergence criterion $\delta$.
		\ENSURE Estimated coefficient $\beta$.
		
		\STATE Initialize $\beta^{1} \leftarrow \beta^0 - \nabla R_{n,c}(\beta^0)$
		\STATE Set iteration counter $t \leftarrow 0$
		
		\REPEAT
			\STATE $t \leftarrow t + 1$
			\STATE Compute difference: $\delta^t \leftarrow \beta^t - \beta^{t-1}$
			\STATE Compute gradient difference: $g^t \leftarrow \nabla R_{n,c}(\beta^t) - \nabla R_{n,c}(\beta^{t-1})$
			
			\STATE Calculate step sizes:
			\STATE \quad $\eta_{1,t} \leftarrow \frac{\langle \delta^t, \delta^t \rangle}{\langle \delta^t, g^t \rangle}$ and $\eta_{2,t} \leftarrow \frac{\langle \delta^t, g^t \rangle}{\langle g^t, g^t \rangle}$
			
			\IF{$\eta_{1,t} > 0$}
				\STATE $\eta_t \leftarrow \min\{\eta_{1,t}, \eta_{2,t}, 100\}$
			\ELSE
				\STATE $\eta_t \leftarrow 1$
			\ENDIF
			
			\STATE Update parameter: $\beta^{t+1} \leftarrow \beta^t - \eta_{t} \cdot \nabla R_{n,c}(\beta^t)$
			
		\UNTIL{$\|\nabla R_{n,c}(\beta^{t})\|_2 < \delta$}
	\end{algorithmic}  
\end{algorithm}

\indent Before applying gradient descent, we standardize the covariates to have zero mean and unit variance. 

\begin{remark}
\indent The computational cost of the proposed method is primarily dictated by the gradient evaluation of the objective function, $\nabla R_{n,c}(\beta)$, within each iteration of the Barzilai-Borwin (BB) algorithm. A key advantage of our approach lies in its computational efficiency. The derivative of our smoothed loss function, $\rho_{\tau,c}^{\prime}(x)$, is composed solely of basic algebraic operations (addition, multiplication, division, and square root), which are executed rapidly on modern hardware. In contrast, prevalent convolution-based methods \cite{He2020SmoothedQR} often yield gradients involving computationally expensive special functions. For instance, Gaussian kernel smoothing results in a derivative $l_h^{\prime}(x) = \tau - \Phi(-\frac{x}{h})$ that requires evaluating the standard normal CDF, $\Phi(x)$, while a logistic kernel involves the exponential function. The evaluation of these transcendental functions relies on numerical approximations and is substantially more costly than simple algebraic operations. This theoretical computational advantage is empirically confirmed in Figure 3, which illustrates the superior speed of our method.
\end{remark}

 \begin{figure*} 
  \centering
    {\includegraphics[width = 0.6\textwidth]{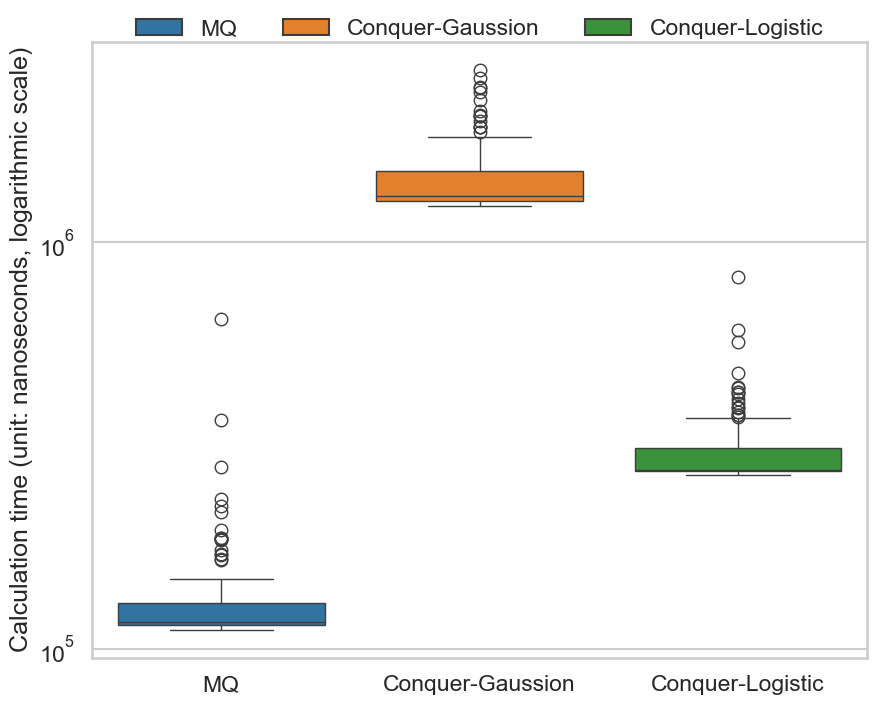}}
  \caption{Comparison of computation time for first derivative of different loss functions.}
  \label{figFunction_time_comparison}
\end{figure*}

\begin{remark}
\indent Beyond per-iteration efficiency, a key advantage of our MQ-based smoothing lies in the quality of the second-order information it provides to the Barzilai-Borwein (BB) algorithm. Since the BB step length implicitly approximates the inverse of the Hessian, the behavior of the objective's second derivative is critical. Our method's second derivative, $\rho_{\tau,c}''(x)$, exhibits slow algebraic decay ($O(|x|^{-3})$), whereas the second derivative of convolution-based methods (e.g., Gaussian kernel) decays exponentially. This distinction is crucial during optimization. The exponential decay effectively nullifies the contribution of data points with large residuals to the Hessian approximation, meaning the curvature estimate is dominated by already well-fitted points. In contrast, the slower algebraic decay of our method ensures that all data points, even those with large errors, contribute meaningfully to the curvature estimate. Consequently, the BB algorithm is informed by a more global and robust curvature, leading to more appropriate step lengths and a more efficient convergence trajectory.
\end{remark}

 \begin{figure*} 
  \centering
    {\includegraphics[width = 0.6\textwidth]{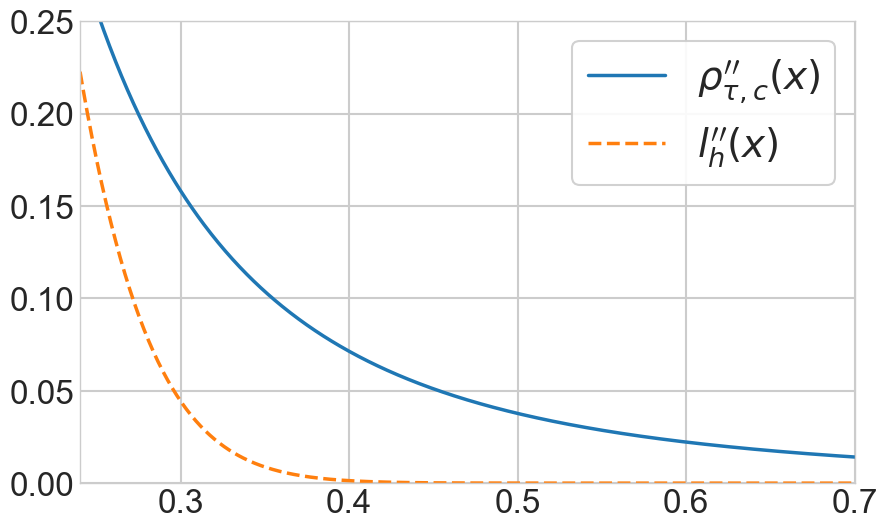}}
  \caption{Comparison of second derivative value of different loss functions.}
  \label{secondDerivativeValueOfMQAndConvol}
\end{figure*}

\indent In summary, while both smoothing strategies provide the necessary differentiability to apply gradient-based algorithms, their per-iteration computational costs differ substantially. The gradient evaluation for the MQ-based smoothing strategy relies entirely on computationally inexpensive algebraic operations. In contrast, convolution-based kernel smoothing introduces computationally intensive transcendental functions (e.g., the CDF or the exponential function). Consequently, under identical hardware conditions, the MQ-smoothed loss function results in a lower wall-clock time per iteration. This efficiency advantage becomes particularly significant for large-scale datasets, where the cumulative time saved over millions of gradient evaluations can be substantial.

\newpage
\numberwithin{equation}{section}
\setcounter{equation}{0}
\section{Numerical Simulation }\label{section:NumericalSimulation}
\indent In this section, we use a loss function based on MQ function for numerical simulation to verify the smoothing effect of the loss function. We mainly focus on linear quantile regression and its related regression models
\subsection{Quantile regression}\label{subsection:NumSim_QR}
In this section, we apply the proposed generalized MQ function to smooth the loss function in quantile regression. Considering real-world problems, especially in today's internet age, the amount of data is increasing day by day. In performing regression analysis, the raw data we can use is also approaching the limit of computer memory storage. Therefore, for internet data, we can usually obtain sufficient data so that many of the most primitive data analysis methods can no longer handle such massive data. Thus, we consider large sample sizes and examine the experimental effects of our proposed method through numerical simulations.Using the linear quantile regression model ($F_{y|x}^{-1} (\tau)= x^T\beta^* (\tau)$), given the data vectors $(x,y)$ and quantile level $\tau \in(0,1)$, we can write it in the form of a linear model:
\begin{equation}\label{3.1}
    y = x^T\beta^* (\tau) + \epsilon(\tau)
\end{equation}
where the random variable $\epsilon(\tau)$ satisfies $P\{\epsilon(\tau) \leq 0|x\}=\tau$, the random error term follows a Gaussian distribution $\mathcal{N}(0,4)$. We generate the response variable $y_i$ using the following model: 
\begin{equation}\label{3.2}
    y_i=x_i^T\beta^* + \{\varepsilon_i-F_{\varepsilon_i}^{-1}(\tau)\},i=1,2,\ldots,n;
\end{equation}
\indent To evaluate the performance of the methods, we use the $L_2$ norm of the estimation error, i.e., $||\hat{\beta}-\beta^*||_2$, and record the computational time. We compare our proposed generalized MQ function with convolution-based smoothing quantile regression (They refer to it as "Conquer" ) proposed by Xuming He et al. \cite{He2020SmoothedQR}. Using the kernel-based convolution smoothing method involves the choice of kernel function and a smoothing parameter $h$. He et al. \cite{He2020SmoothedQR} illustrates five commonly used kernel functions in their work, and concludes through simulations that the "Gaussian"-based method is the most effective. Therefore, in all our simulation studies using the convolution smoothing method, we take the kernel function as the "Gaussian" and "Logistic" kernel, and the smoothing parameter $h$ as $h = {(p + log n)/n}^{2/5}$, where $n$ is the sample size and $p$ is the number of covariates. The experiments in this section are based on an Intel Core I7-6700 3.4GHz computer with 16GB memory.

\begin{figure*}[htbp]
  \centering

  \begin{subfigure}[t]{0.48\textwidth}
    \centering
    \includegraphics[width=\textwidth]{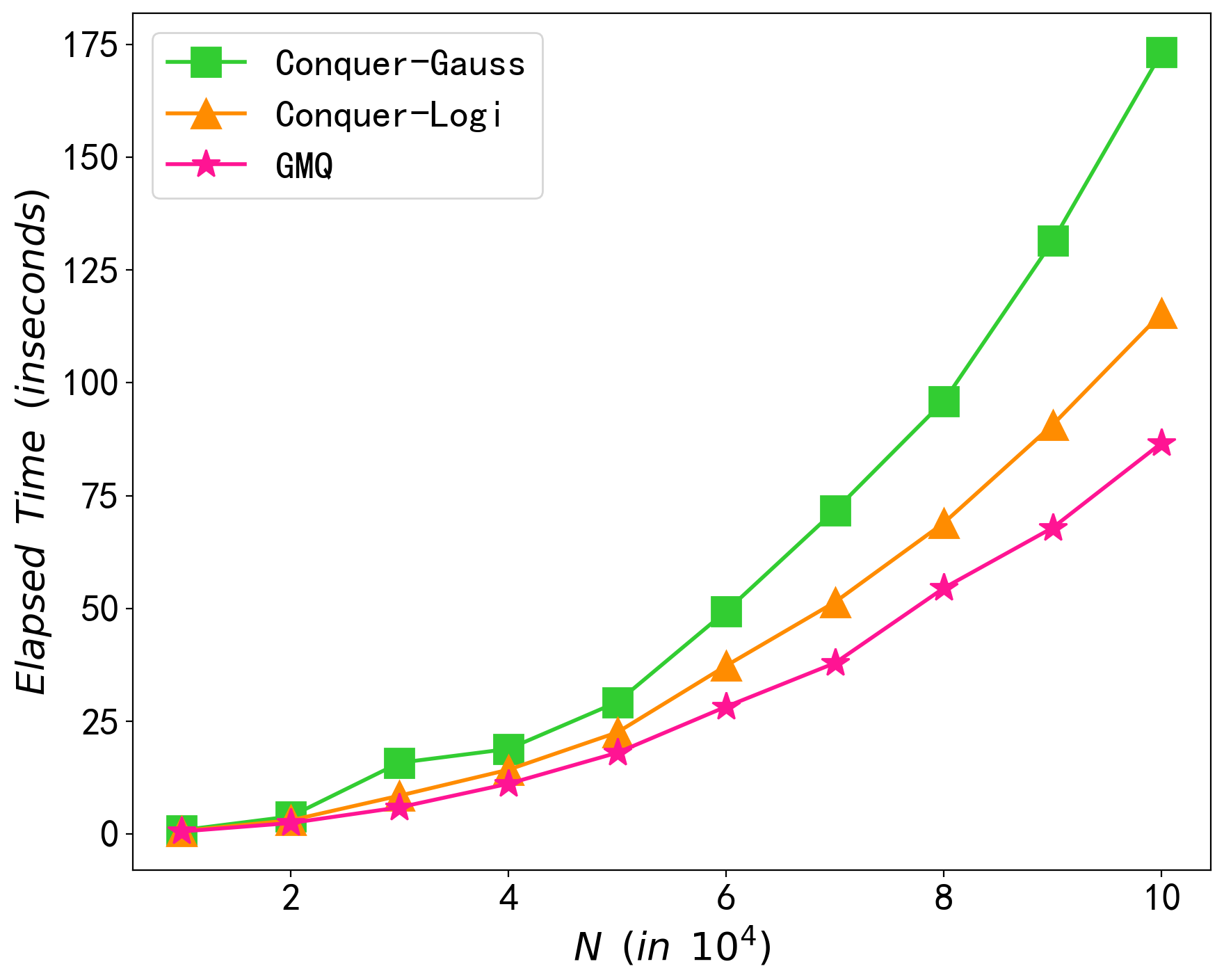}
    \caption{Time consumption with $\mathcal{N}(0,4)$ error.}
    \label{fig4_1}
  \end{subfigure}%
  \hfill
  \begin{subfigure}[t]{0.48\textwidth}
    \centering
    \includegraphics[width=\textwidth]{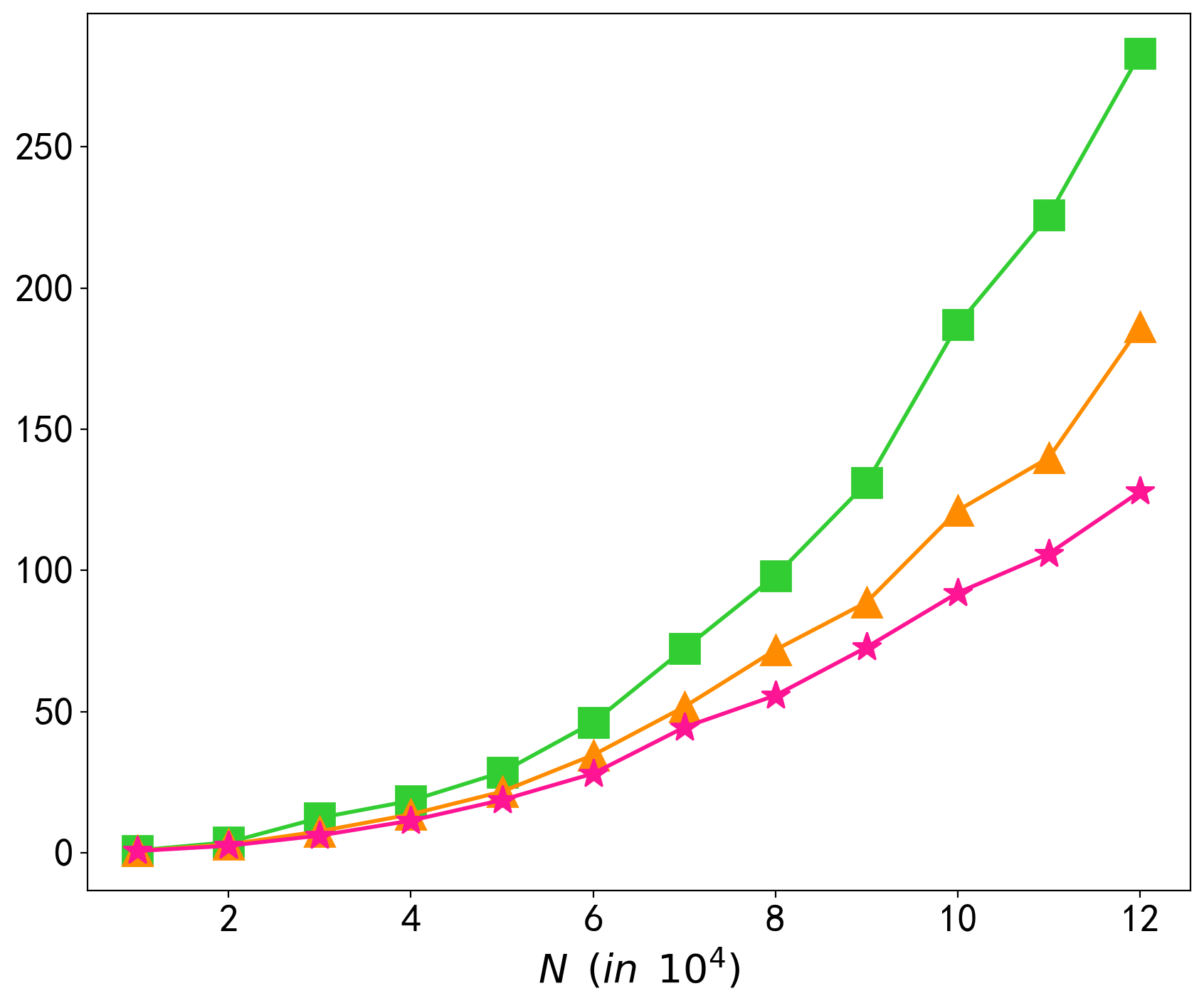}
    \caption{Time consumption with $t_2$ error.}
    \label{fig4_2}
  \end{subfigure}
  
  \vspace{1em} 

  \begin{subfigure}[t]{0.48\textwidth}
    \centering
    \includegraphics[width=\textwidth]{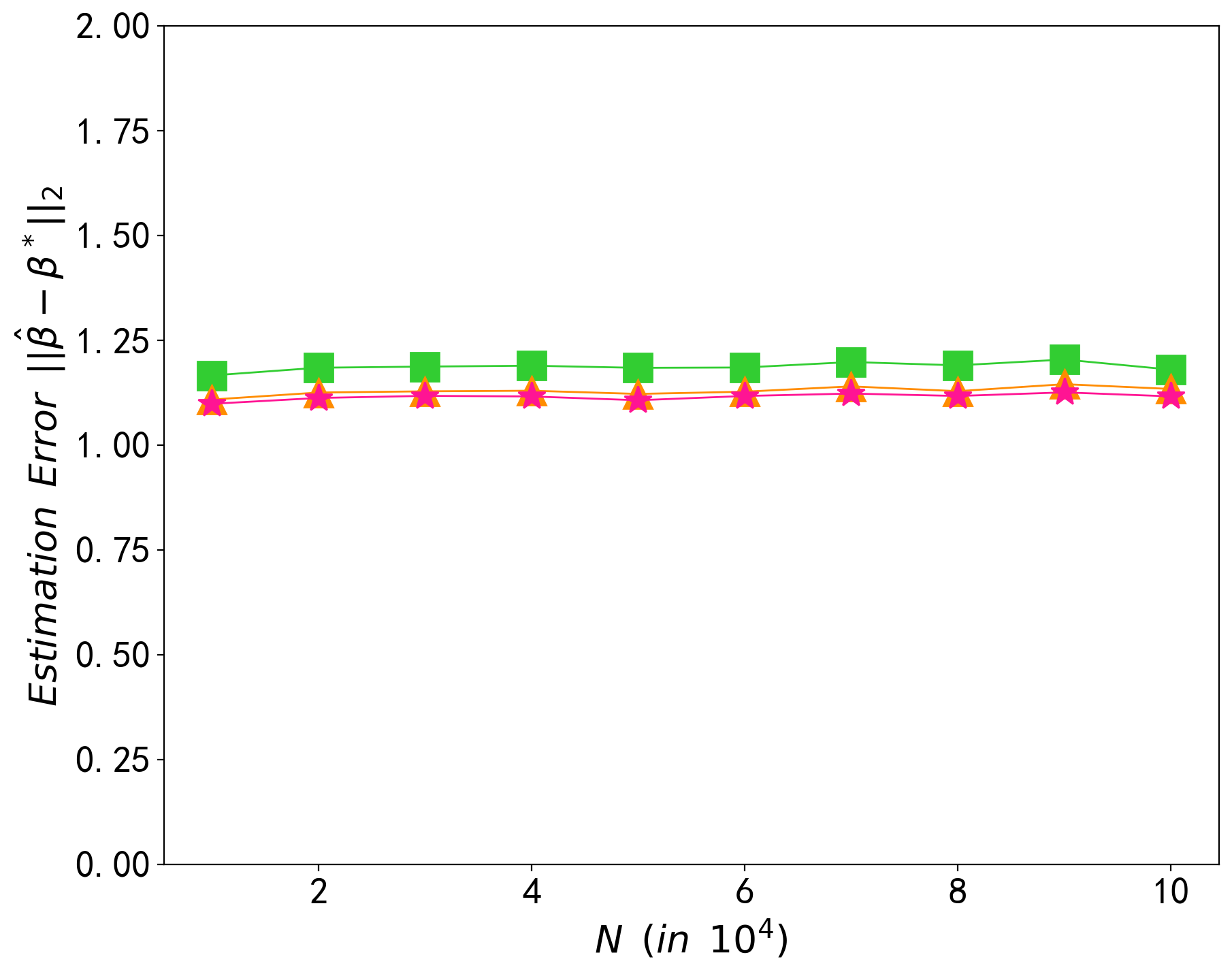}
    \caption{Estimation error with $\mathcal{N}(0,4)$ error.}
    \label{fig4_3}
  \end{subfigure}%
  \hfill
  \begin{subfigure}[t]{0.48\textwidth}
    \centering
    \includegraphics[width=\textwidth]{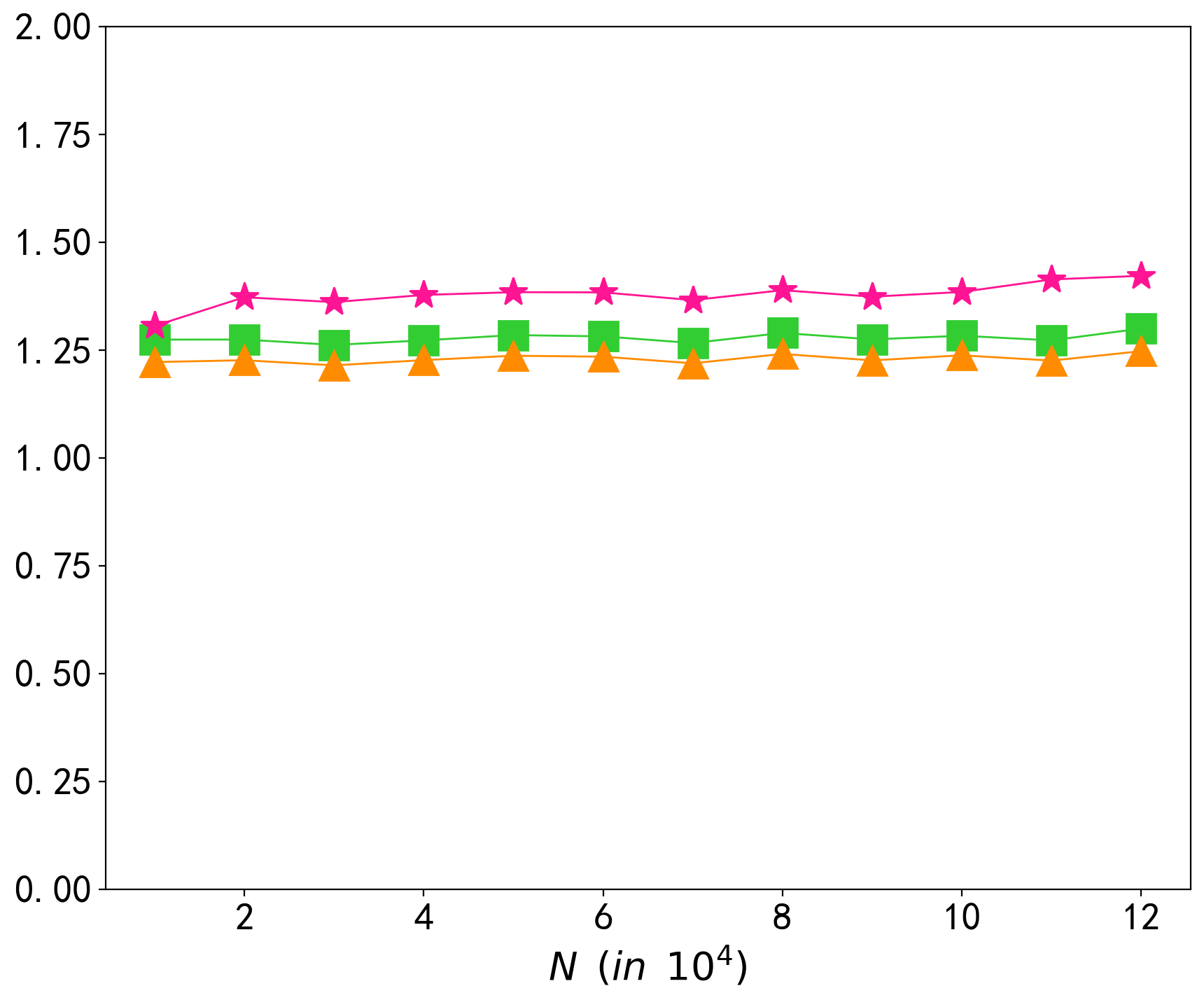}
    \caption{Estimation error with $t_2$ error.}
    \label{fig4_4}
  \end{subfigure}

  \caption{Model (\ref{3.2}) comparison: Time consumption and estimation error based on convolutional smoothing methods (Gaussian and Logistic kernels) versus the GMQ smoothing method.}
  \label{fig4}
\end{figure*}

\indent He et al. \cite{He2020SmoothedQR} has demonstrated advantages in terms of time and errors compared to standard quantile regression for sample sizes within 5000. Therefore, our experiments are geared towards larger sample sizes (greater than 10000). Throughout all experiments, we maintain the relationship between sample size $N$ and the dimension of the predictor variable $p$ as $N/p=20$. When the dimension is large (exceeding 500), convolution smoothing-based quantile regression and GMQ-based smoothed quantile regression exhibit similar regression errors, with differences in model parameter errors around 0.1. When distributing the errors evenly across the coefficients of each predictor variable, these differences can be considered negligible. However, notably, the GMQ smoothing method demonstrates a more significant advantage in terms of time consumption.
\\ \indent It is worth noting that our method only replaces the loss function with a smooth function, and in special cases ($c=0$), our loss function degrades to the traditional quantile regression loss function. 

\subsection{Expectile regression}\label{subsection:ER}
\indent For expectile regression, we utilize the same algorithm as MQ-based smoothed quantile regression for comparison. The standard expectile regression method can be found in the R language package "expectreg." We introduce two models to generate sample data:
\begin{equation}\label{3.3}
    y_i=\beta_0^* + x_i^T\beta^* + (0.5x_{i,p}+1)\{\varepsilon_i-F_{\varepsilon_i}^{-1}(\tau)\},i=1,2,\ldots,n;
\end{equation}
\begin{equation}\label{3.4}
    y_i=\beta_0^* + x_i^T\beta^* + 0.5((x_{i,p}+1)^2+1)\{\varepsilon_i-F_{\varepsilon_i}^{-1}(\tau)\},i=1,2,\ldots,n;
\end{equation}
\\ \indent Random errors are generated from two different distributions: a $t$-distribution with 2 degrees of freedom and a Gaussian distribution  $\mathcal{N}(0,4)$. We conduct experiments comparing parameter estimation using MQ-based smoothed Expectile regression with standard Expectile regression. Simultaneously, we assess time consumption and errors.

\begin{figure*}[htbp]
  \centering

  \begin{subfigure}[t]{0.32\textwidth}
    \centering
    \includegraphics[width=\textwidth]{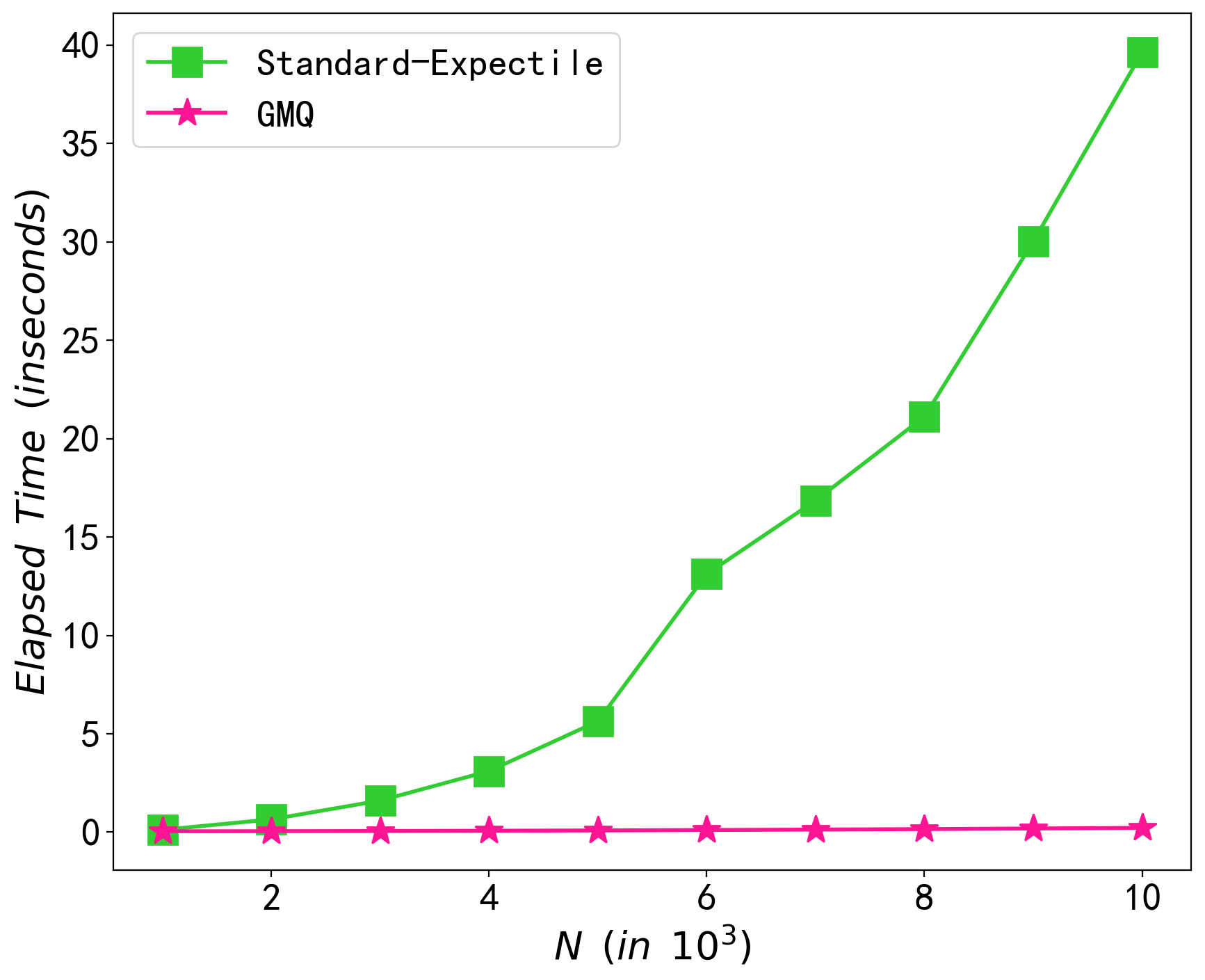}
    \caption{Model (\ref{3.2}) with $\mathcal{N}(0,4)$ error.}
    \label{fig5_1}
  \end{subfigure}%
  \hfill
  \begin{subfigure}[t]{0.32\textwidth}
    \centering
    \includegraphics[width=\textwidth]{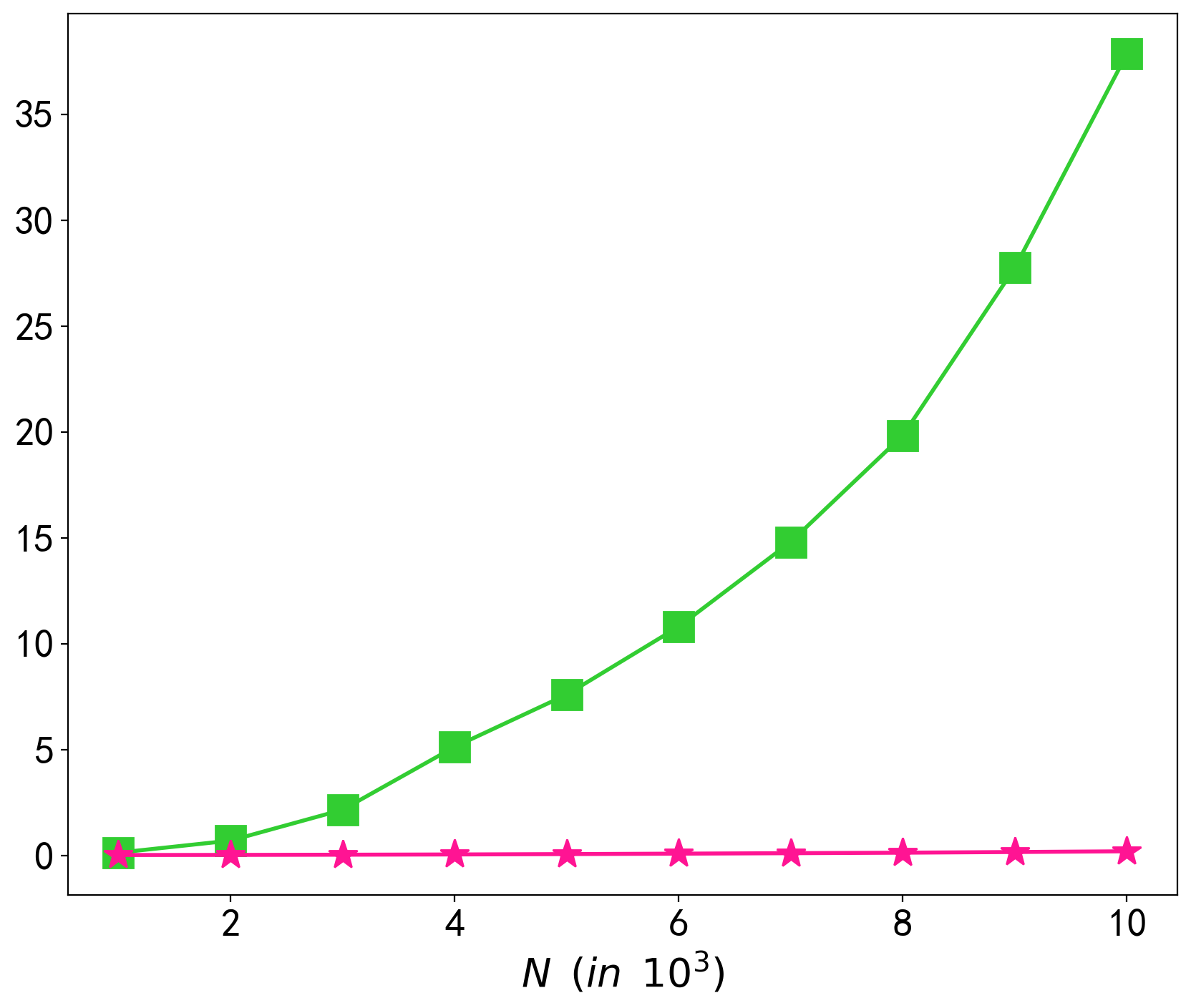}
    \caption{Model (\ref{3.3}) with $\mathcal{N}(0,4)$ error.}
    \label{fig5_2}
  \end{subfigure}%
  \hfill
  \begin{subfigure}[t]{0.32\textwidth}
    \centering
    \includegraphics[width=\textwidth]{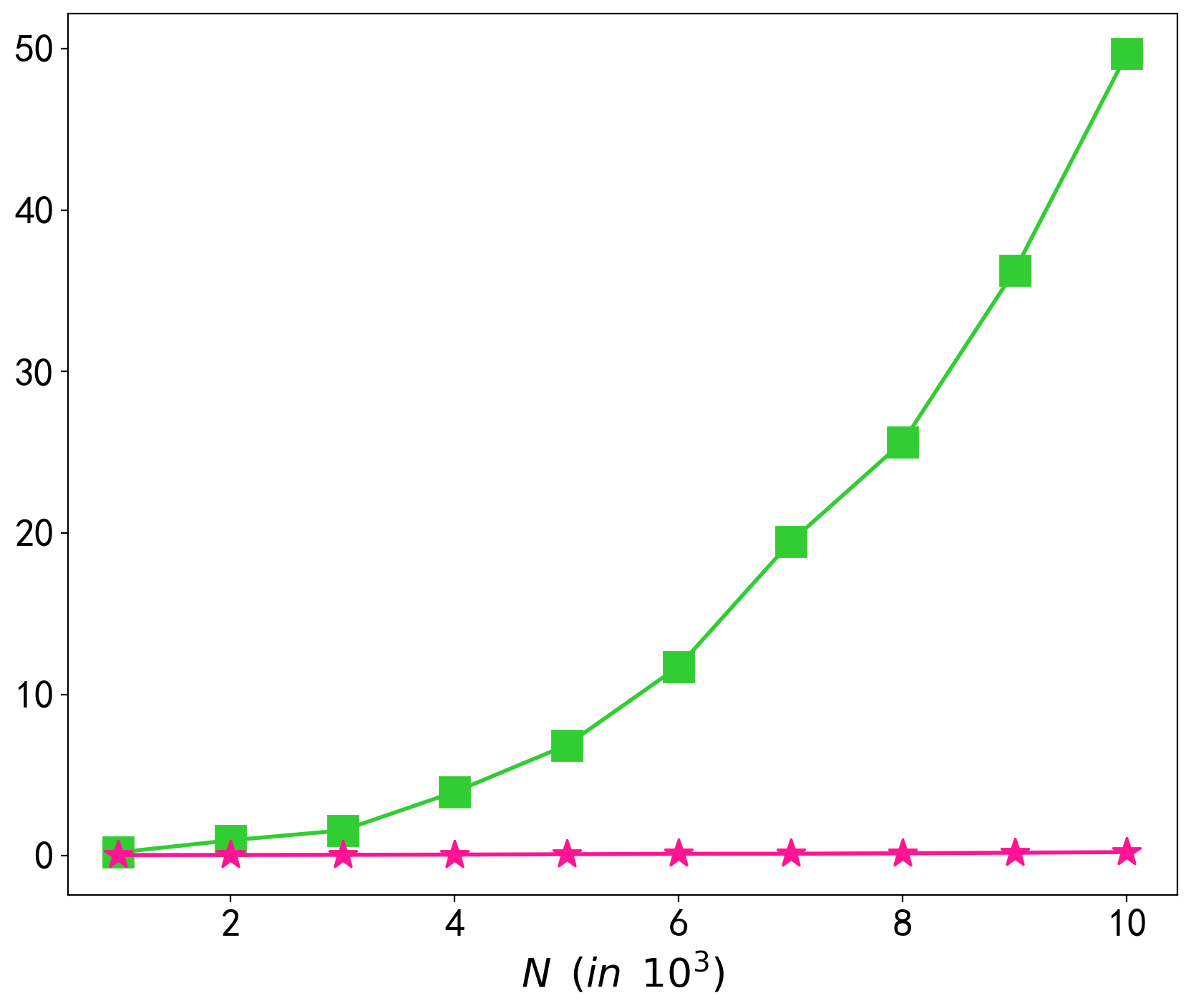}
    \caption{Model (\ref{3.4}) with $\mathcal{N}(0,4)$ error.}
    \label{fig5_3}
  \end{subfigure}

  \vspace{1em} 
  
  \begin{subfigure}[t]{0.32\textwidth}
    \centering
    \includegraphics[width=\textwidth]{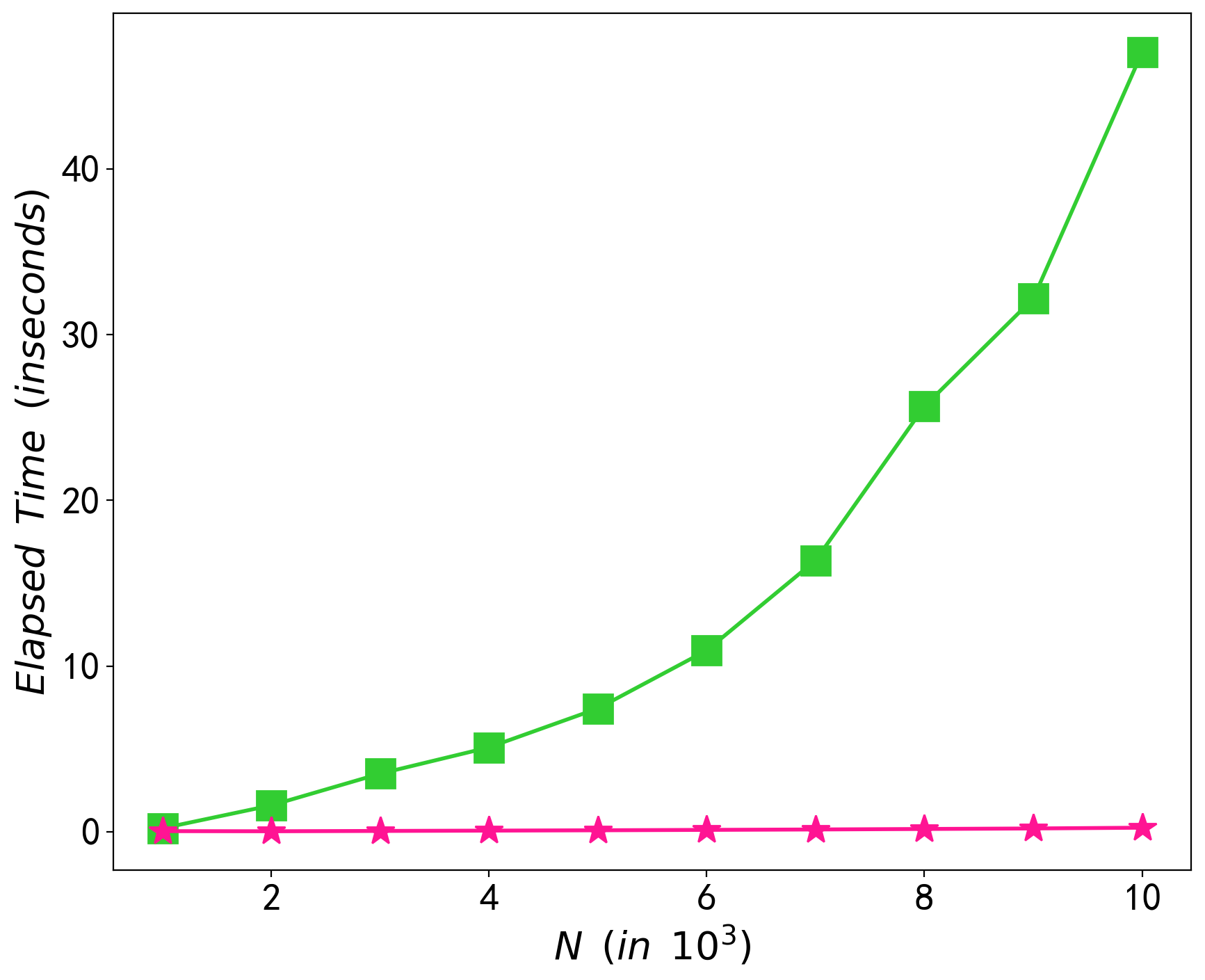}
    \caption{Model (\ref{3.2}) with $t_2$ error.}
    \label{fig5_4}
  \end{subfigure}%
  \hfill
  \begin{subfigure}[t]{0.32\textwidth}
    \centering
    \includegraphics[width=\textwidth]{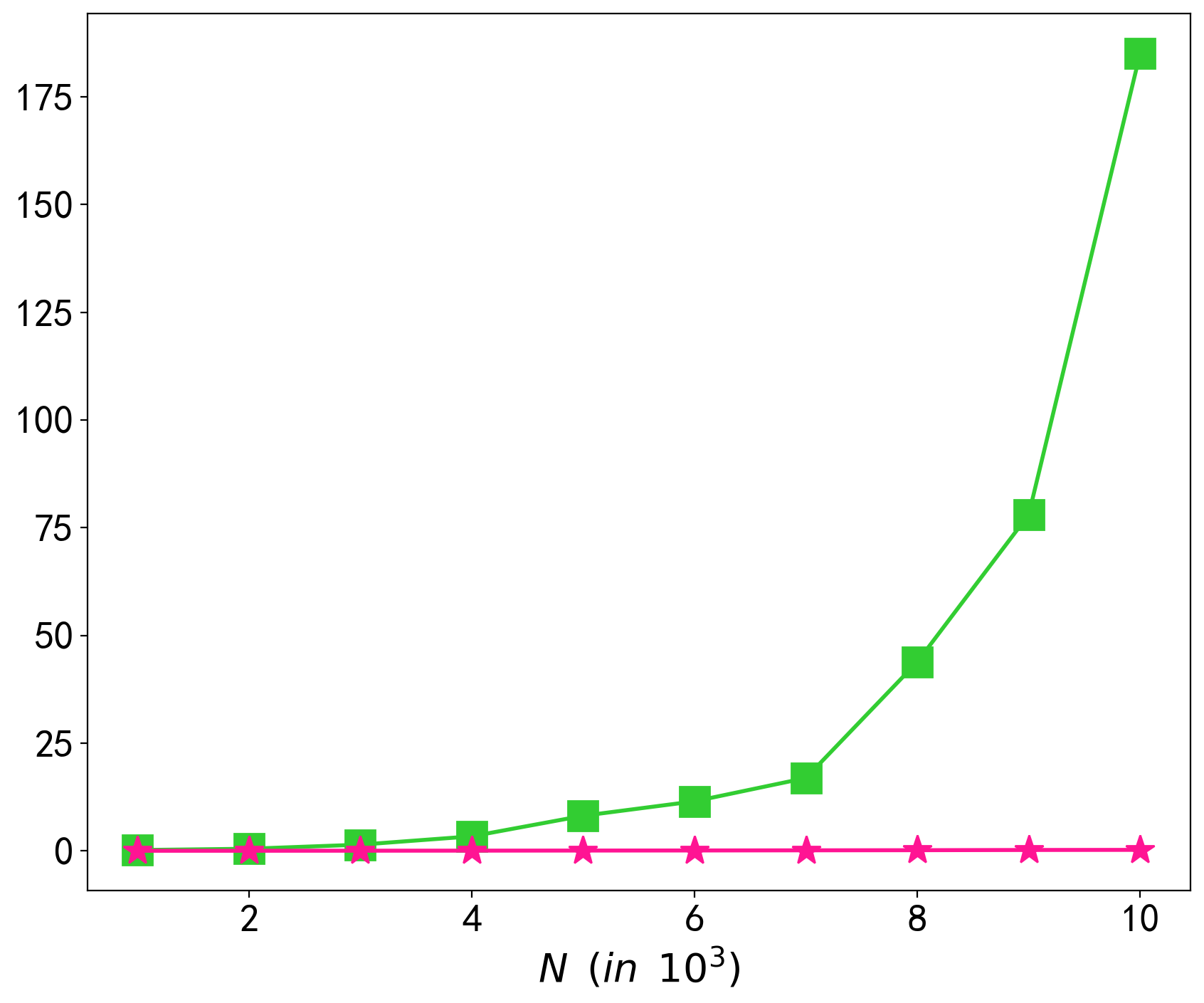}
    \caption{Model (\ref{3.3}) with $t_2$ error.}
    \label{fig5_5}
  \end{subfigure}%
  \hfill
  \begin{subfigure}[t]{0.32\textwidth}
    \centering
    \includegraphics[width=\textwidth]{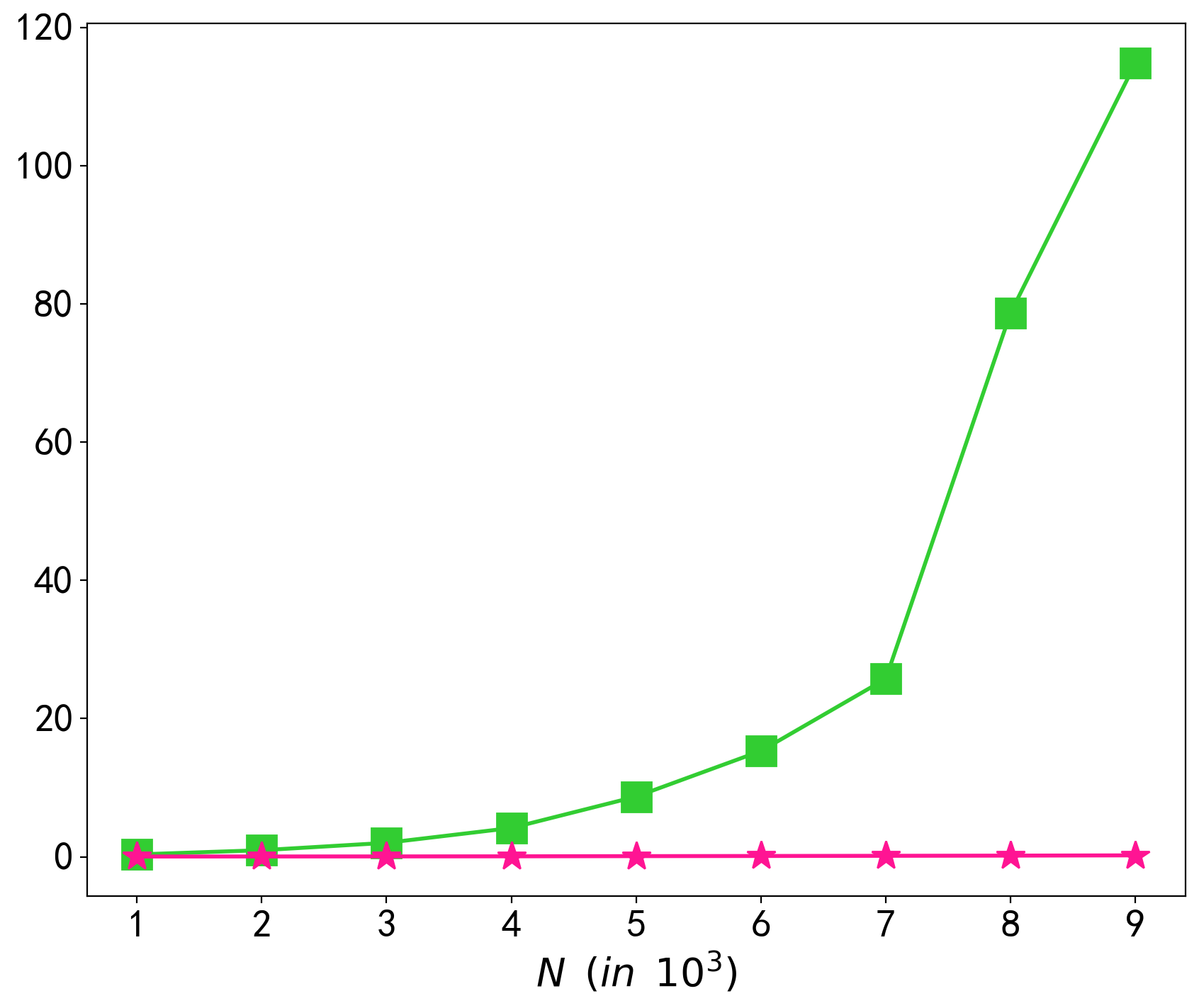}
    \caption{Model (\ref{3.4}) with $t_2$ error.}
    \label{fig5_6}
  \end{subfigure}

  \caption{Comparison of regression time consumption under three different data source models and two random error terms.}
  \label{fig5}
\end{figure*}

\begin{figure*}[htbp]
  \centering
  
  \begin{subfigure}[t]{0.32\textwidth}
    \centering
    \includegraphics[width=\textwidth]{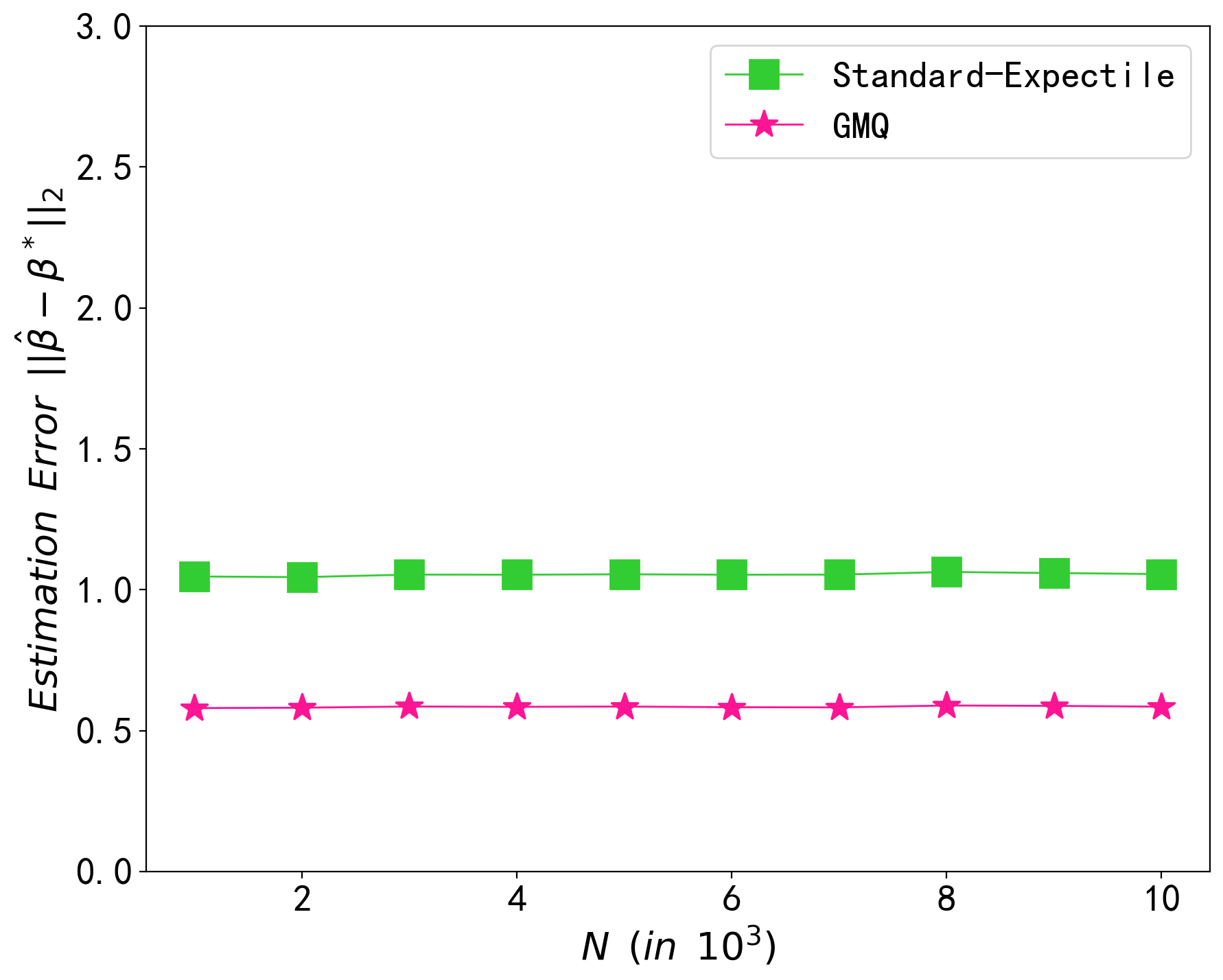}
    \caption{Model (\ref{3.2}) with $\mathcal{N}(0,4)$ error.}
    \label{fig6_1}
  \end{subfigure}%
  \hfill
  \begin{subfigure}[t]{0.32\textwidth}
    \centering
    \includegraphics[width=\textwidth]{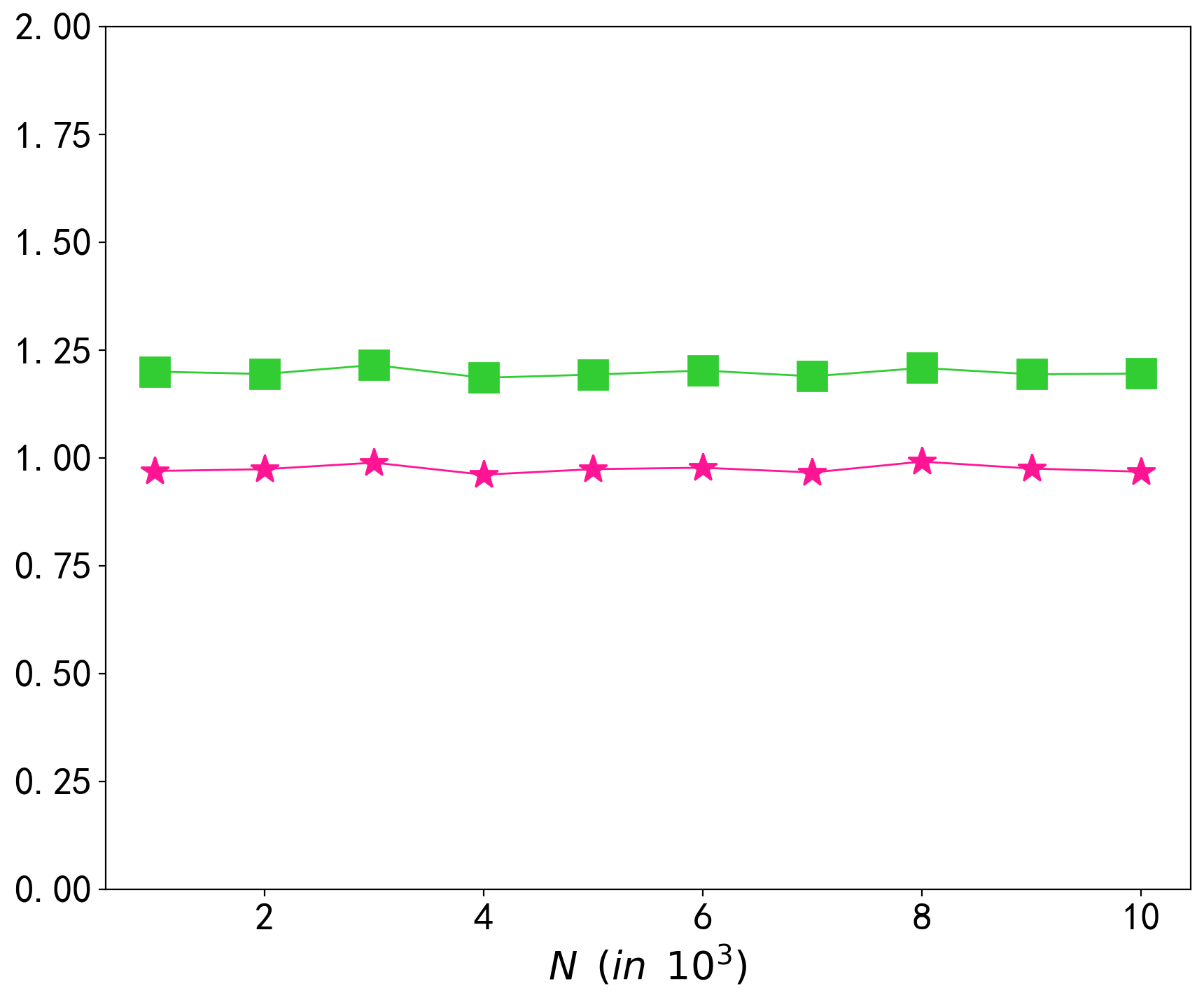}
    \caption{Model (\ref{3.3}) with $\mathcal{N}(0,4)$ error.}
    \label{fig6_2}
  \end{subfigure}%
  \hfill
  \begin{subfigure}[t]{0.32\textwidth}
    \centering
    \includegraphics[width=\textwidth]{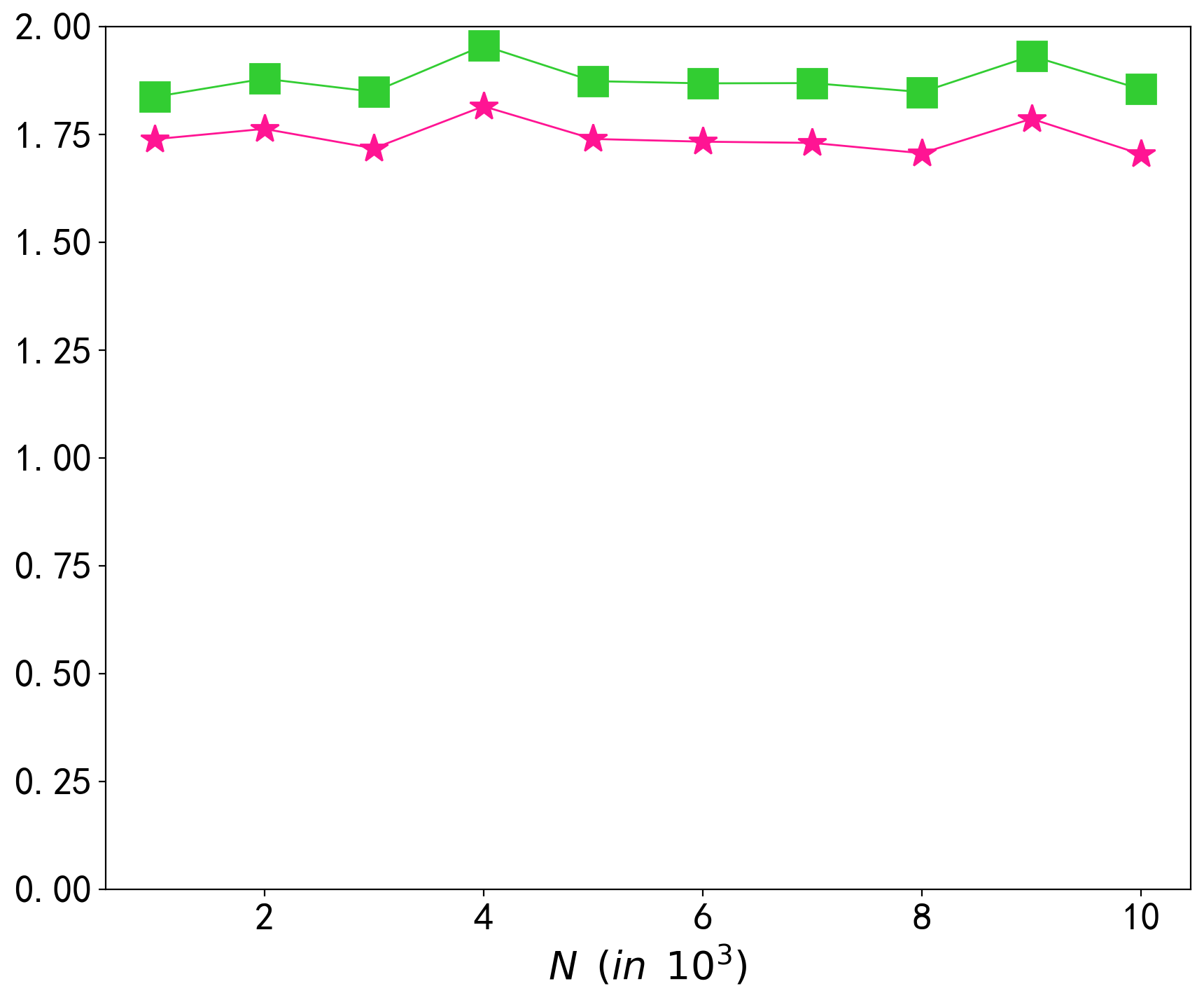}
    \caption{Model (\ref{3.4}) with $\mathcal{N}(0,4)$ error.}
    \label{fig6_3}
  \end{subfigure}

  \vspace{1em} 

  \begin{subfigure}[t]{0.32\textwidth}
    \centering
    \includegraphics[width=\textwidth]{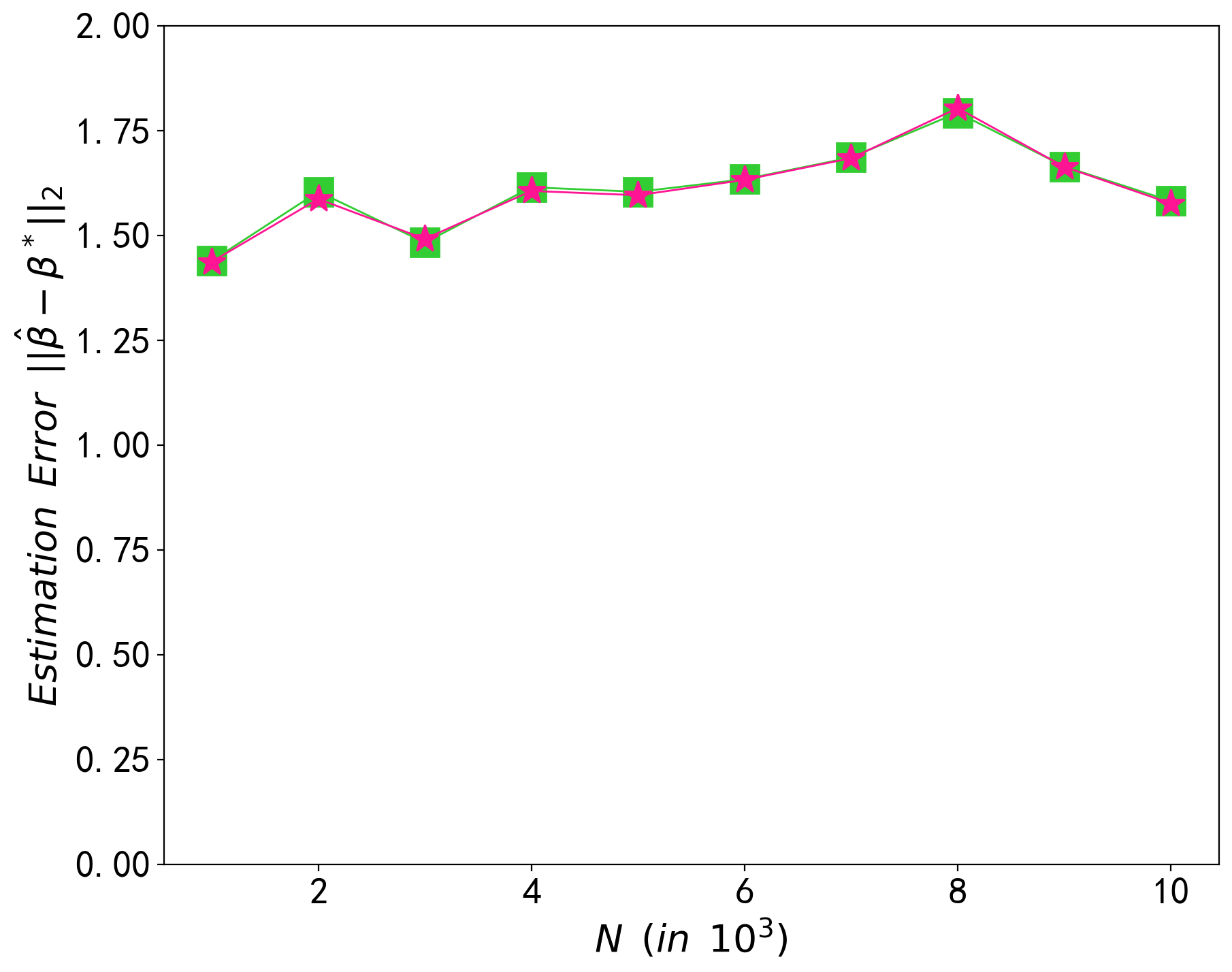}
    \caption{Model (\ref{3.2}) with $t_2$ error.}
    \label{fig6_4}
  \end{subfigure}%
  \hfill
  \begin{subfigure}[t]{0.32\textwidth}
    \centering
    \includegraphics[width=\textwidth]{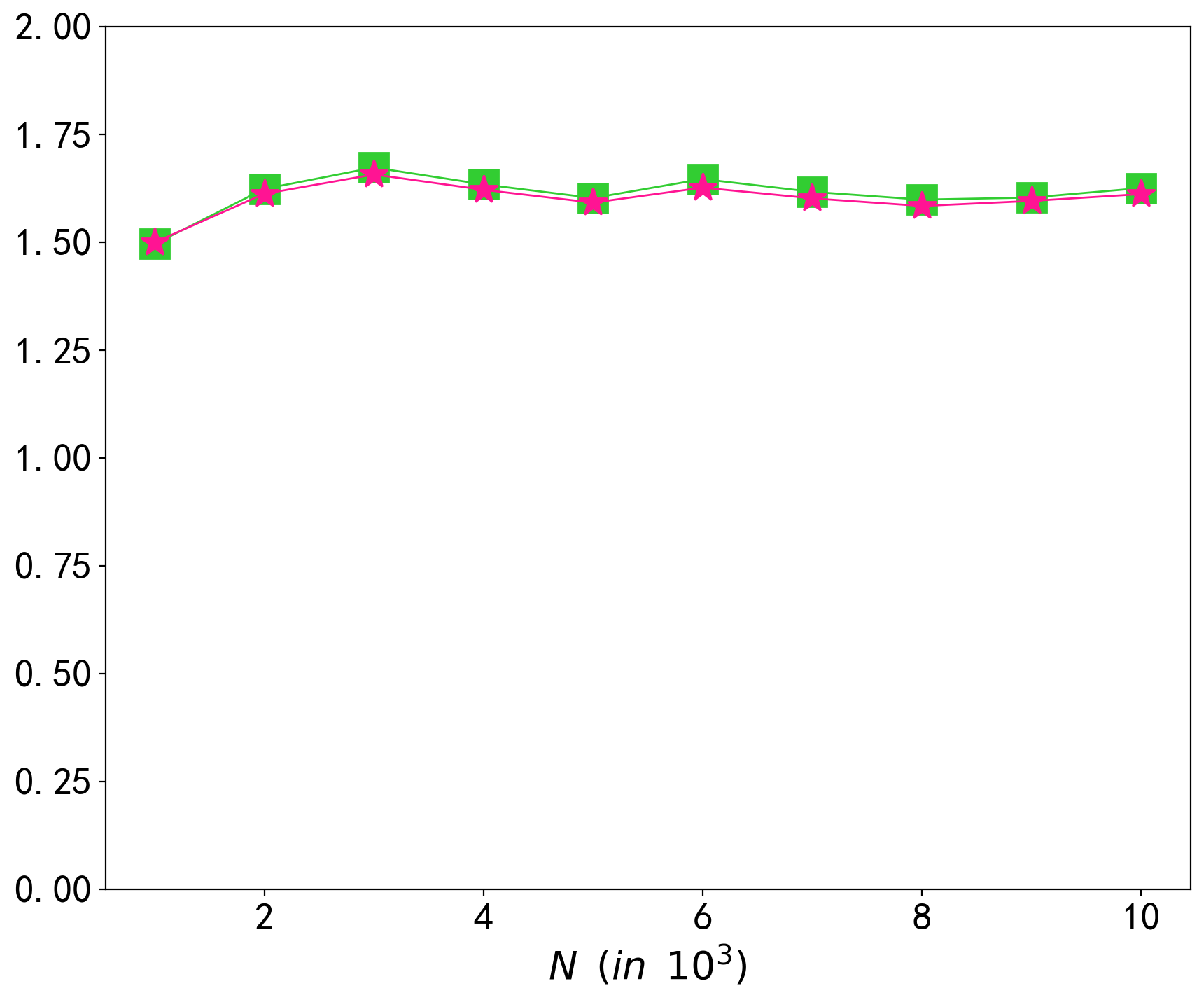}
    \caption{Model (\ref{3.3}) with $t_2$ error.}
    \label{fig6_5}
  \end{subfigure}%
  \hfill
  \begin{subfigure}[t]{0.32\textwidth}
    \centering
    \includegraphics[width=\textwidth]{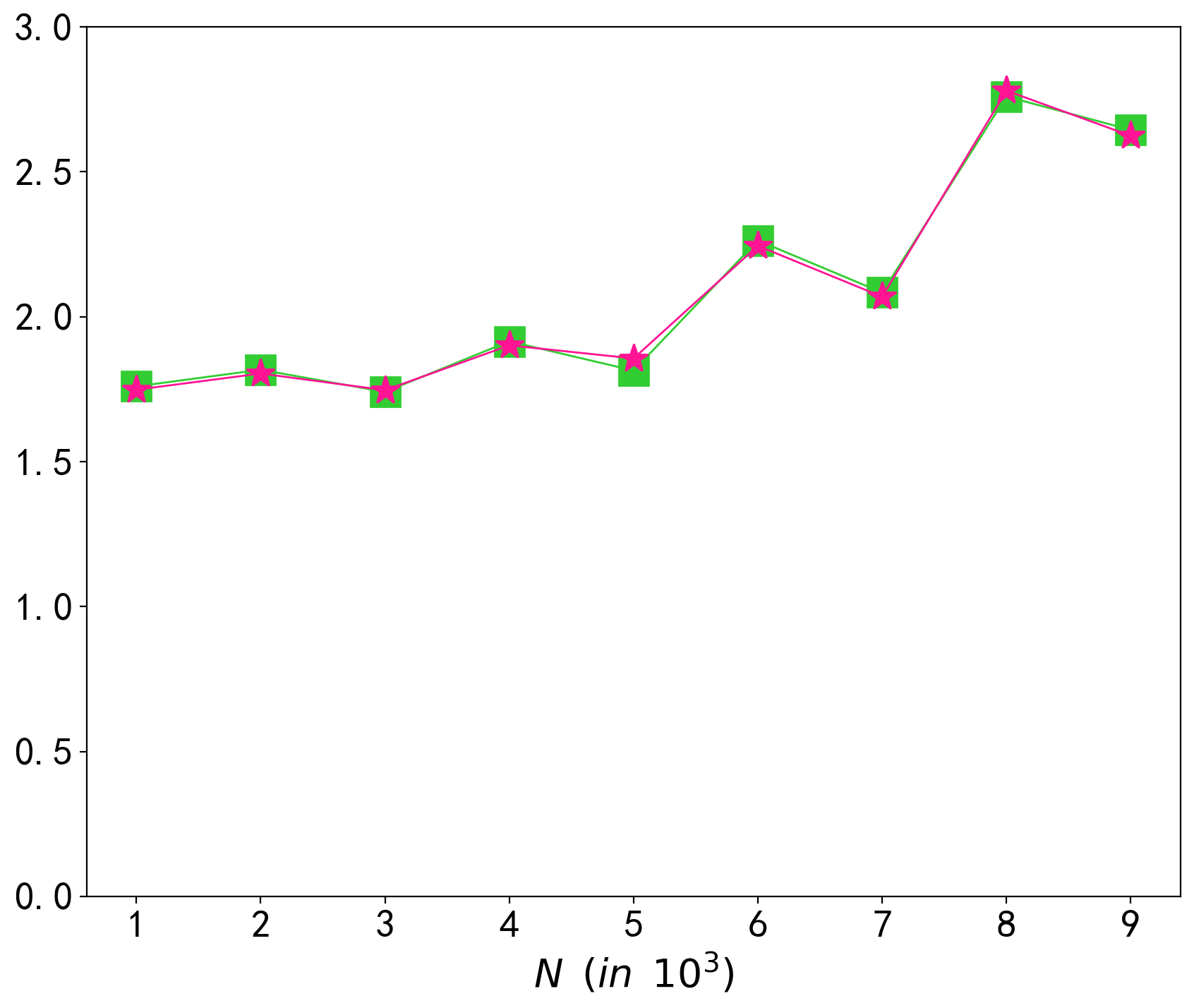}
    \caption{Model (\ref{3.4}) with $t_2$ error.}
    \label{fig6_6}
  \end{subfigure}

  \caption{Comparison of errors under three different data source models and two random error terms.}
  \label{fig6}
\end{figure*}

\indent Figure \ref{fig5} illustrates the time consumption of regression estimates in three different scenarios when $\tau=0.9$. It is observed that, regardless of the data generation model and the distribution of the error term, the smooth Expectile regression constructed based on the MQ-based smoothing method is more efficient in terms of time compared to the standard Expectile regression. As the sample size and dimension increase, the fitting time of the standard Expectile regression sharply increases, while the computational time of the MQ-based function exhibits minimal changes, rendering it negligible compared to the standard Expectile regression.
\\ \indent In Figure \ref{fig6}, the estimation errors of various methods are presented under different simulation conditions when $\tau=0.9$. Across three different models, when the error term follows the $\mathcal{N}(0,4)$ distribution, the estimates from MQ-based outperform the standard Expectile regression. In the case of the error term following a $t$-distribution (with 2 degrees of freedom), the differences between the two methods are not substantial. Although the convolution smoothing method can also be applied to Expectile regression, its implementation ultimately yields results similar to the standard quantile regression. Therefore, a detailed comparison is omitted here.

\subsection{$k$th power expectile regression}\label{subsection:kth}
\indent Finally, we conducted regression experiments on the smooth $kth$ power expectile regression loss functions, with $k$ values chosen as 4/3, 5/3, and 3/2, and sample sizes ranging from 1000 to 5000. We utilized the MQ-based function-based smooth $kth$ power Expectile regression loss function for regression fitting, providing information on fitting time and error rates.

\begin{figure*}[htbp]
  \centering
  
  \begin{subfigure}[t]{0.32\textwidth}
    \centering
    \includegraphics[width=\textwidth]{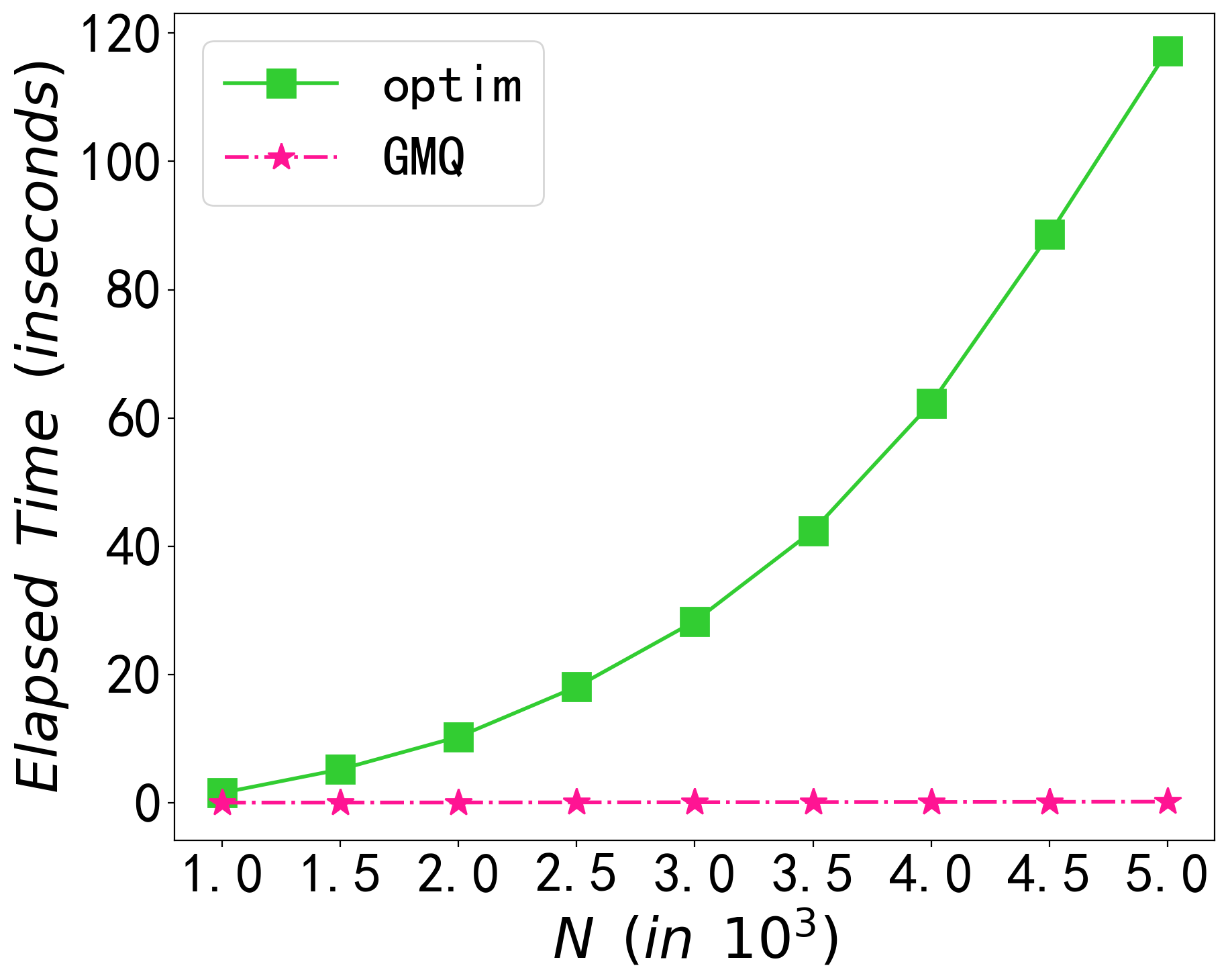}
    \caption{Time consumption with $k=5/3$.}
    \label{fig7_1}
  \end{subfigure}%
  \hfill
  \begin{subfigure}[t]{0.32\textwidth}
    \centering
    \includegraphics[width=\textwidth]{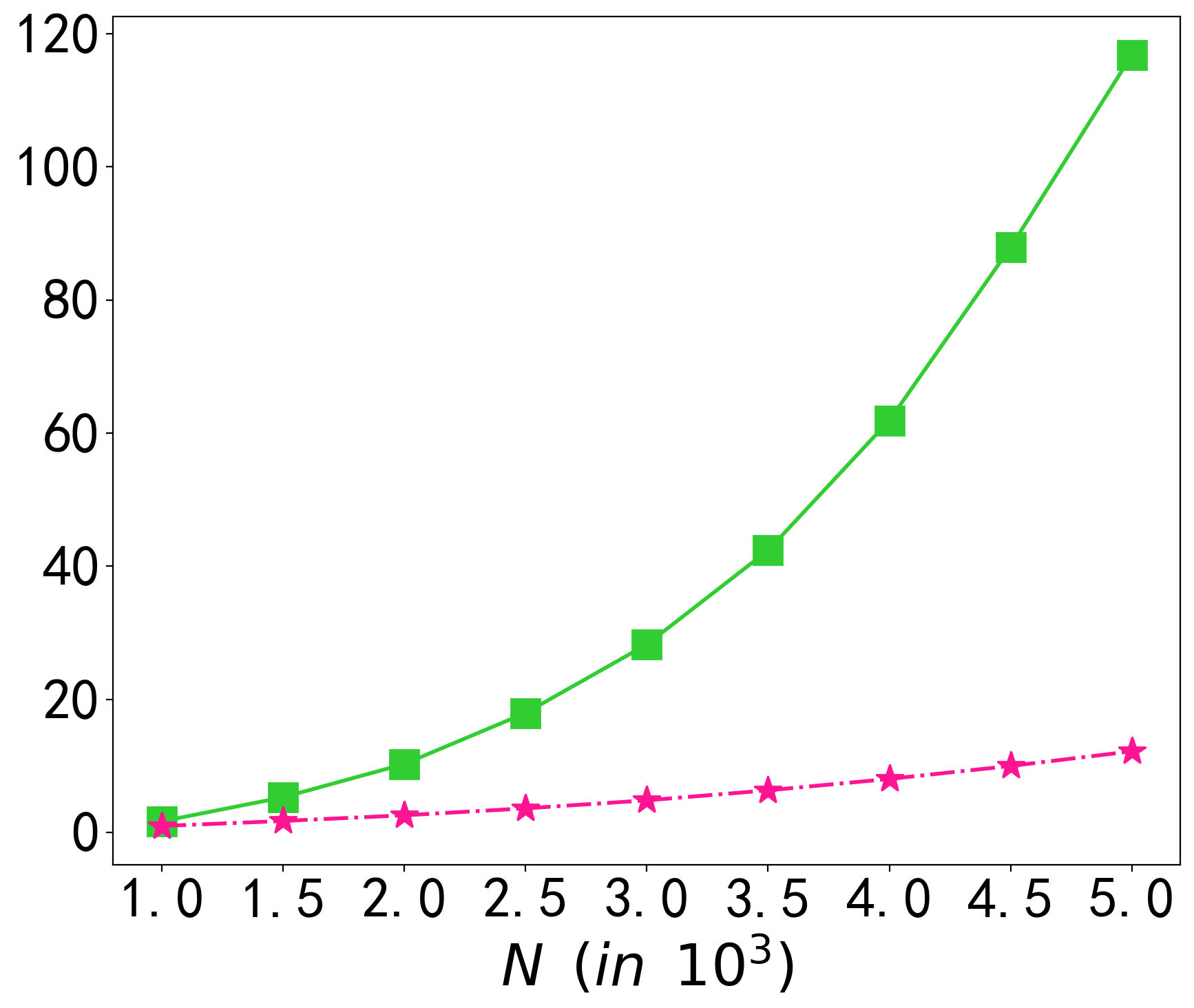}
    \caption{Time consumption with $k=4/3$.}
    \label{fig7_2}
  \end{subfigure}%
  \hfill
  \begin{subfigure}[t]{0.32\textwidth}
    \centering
    \includegraphics[width=\textwidth]{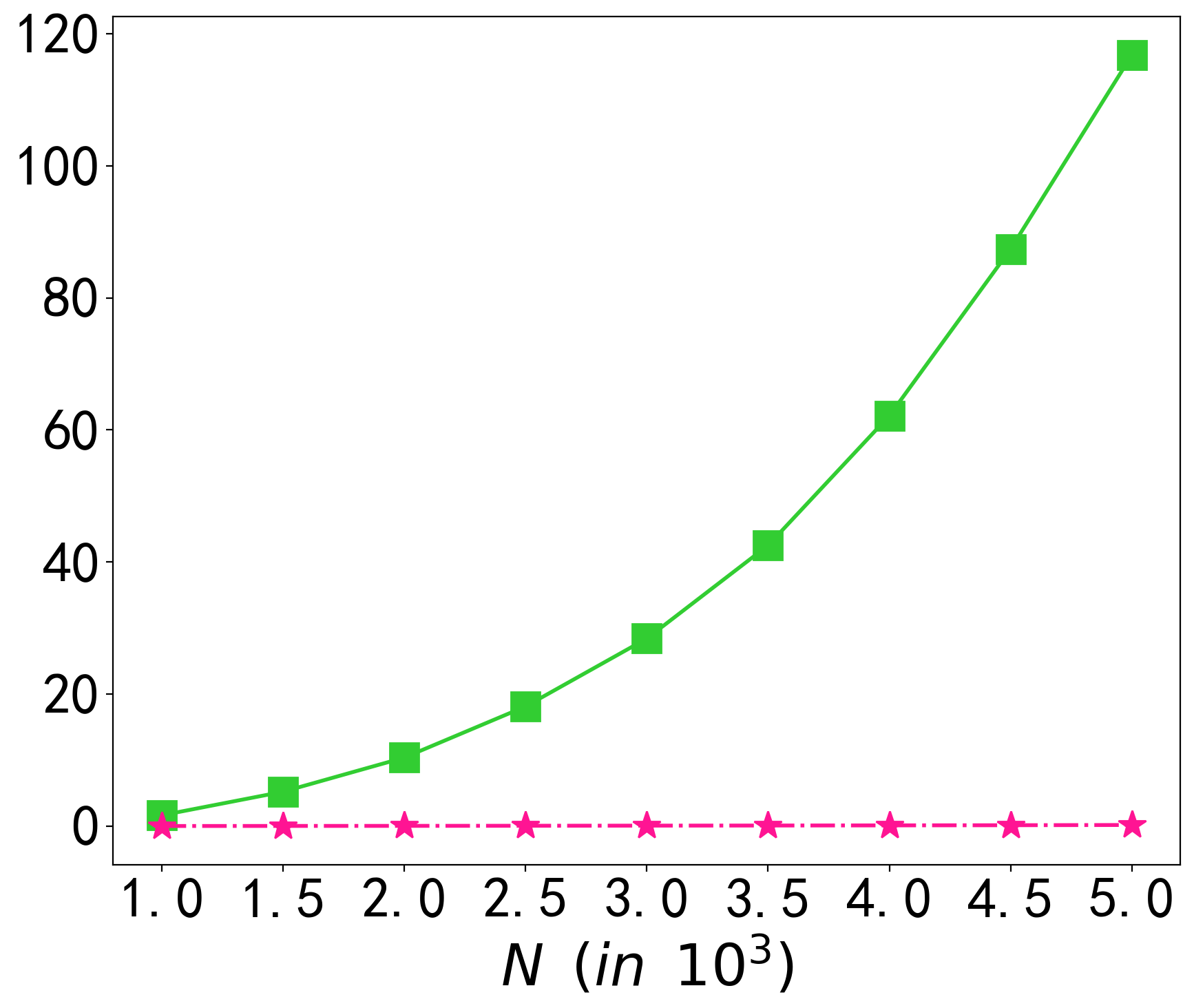}
    \caption{Time consumption with $k=3/2$.}
    \label{fig7_3}
  \end{subfigure}

  \vspace{1em}

  \begin{subfigure}[t]{0.32\textwidth}
    \centering
    \includegraphics[width=\textwidth]{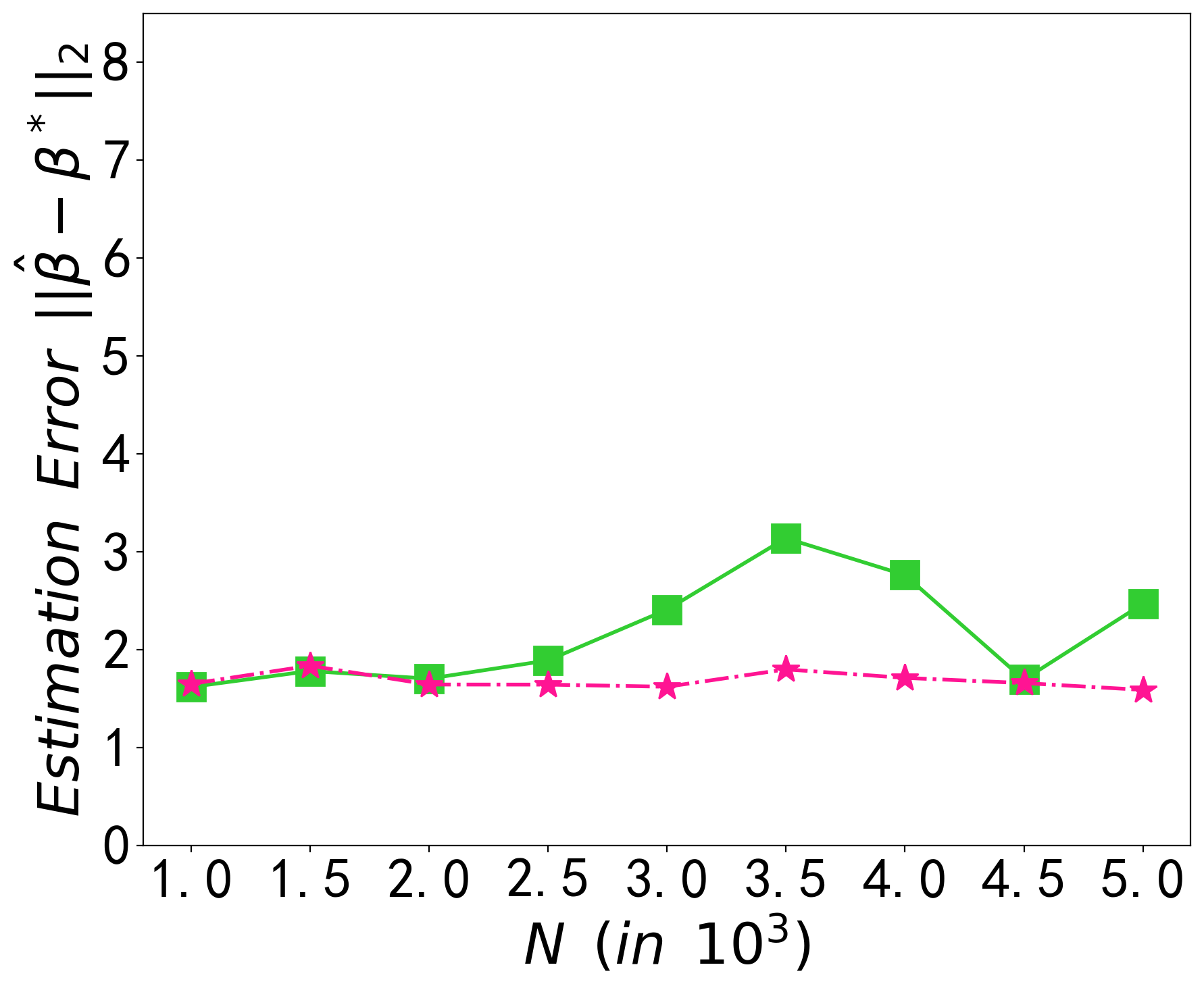}
    \caption{Estimation error with $k=5/3$.}
    \label{fig7_4}
  \end{subfigure}%
  \hfill
  \begin{subfigure}[t]{0.32\textwidth}
    \centering
    \includegraphics[width=\textwidth]{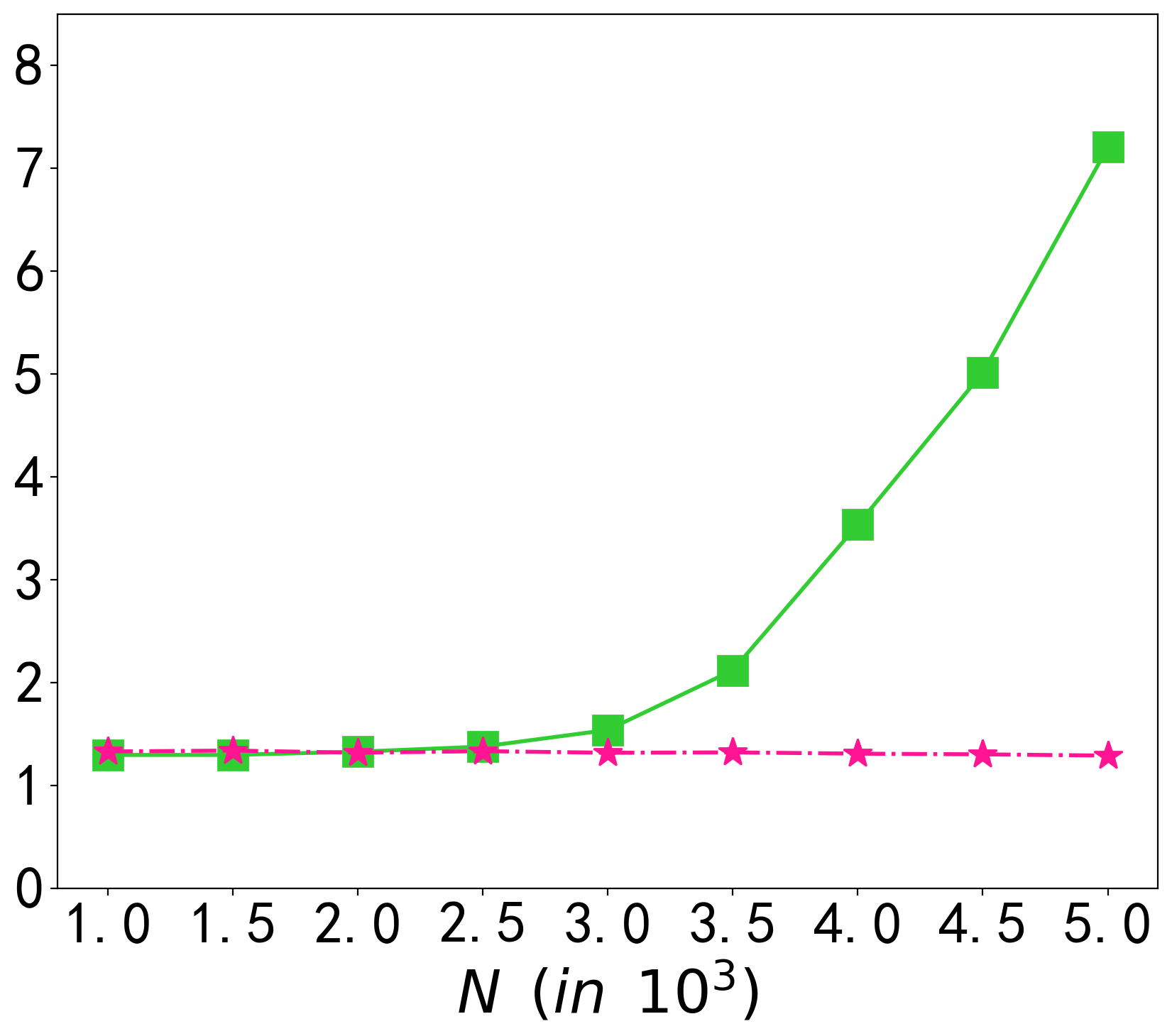}
    \caption{Estimation error with $k=4/3$.}
    \label{fig7_5}
  \end{subfigure}%
  \hfill
  \begin{subfigure}[t]{0.32\textwidth}
    \centering
    \includegraphics[width=\textwidth]{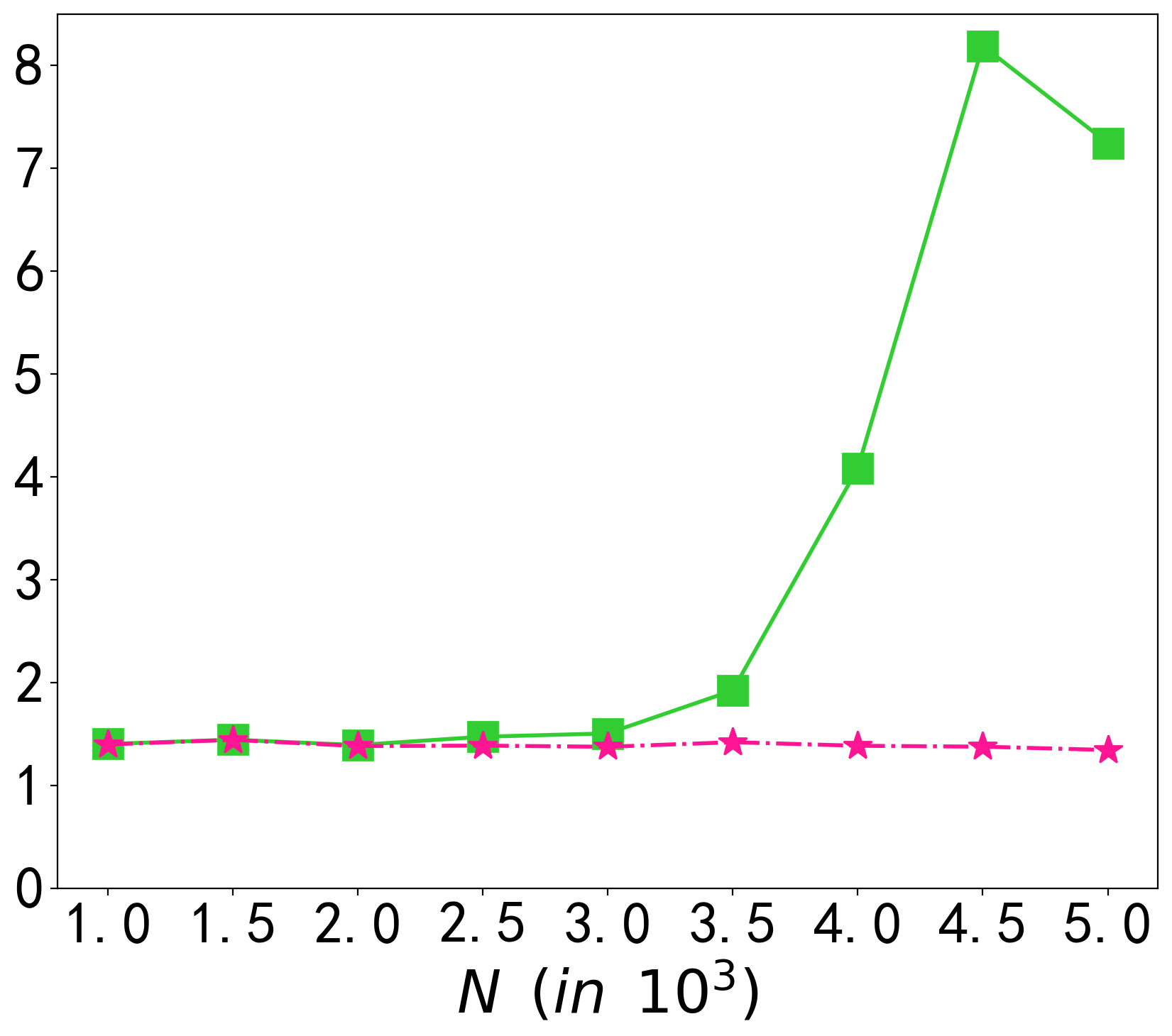}
    \caption{Estimation error with $k=3/2$.}
    \label{fig7_6}
  \end{subfigure}

  \caption{Comparison of regression time and estimation error under model (\ref{3.2}) for parameter values $k = 5/3, 4/3, \text{and } 3/2$.}
  \label{fig7}
\end{figure*}

\indent Figure \ref{fig7} presents experimental results indicating that, when $\tau=0.9$, using the MQ-based function to smooth loss functions for $k=$5/3 and $k=$3/2 yields favorable regression coefficient estimates with minimal computational time. For $k=$4/3, as sample size and dimension increase, the computation time also rises rapidly, but the error remains acceptable and stable.

\newpage
\numberwithin{equation}{section}
\setcounter{equation}{0}
\section{Conclusion}
  We address  computational challenges inherent in standard quantile regression due to the non-differentiable check loss function by proposing  a novel smoothing technique based on GMQ function.  Unlike prevalent convolution-based smoothing techniques, which heavily rely on kernel selection and often lack intuitive interpretation, our technique offers a clear geometric interpretation by constructing the smooth loss as a hyperbola approximating the absolute value function. We establish theoretical  error bounds and asymptotic properties for GMQ-based estimator. 

Our GMQ-based smoothing technique is not limited to quantile regression but can be extended to  $k$th power expectile regression for any $1 \le k \le 2$. While convolution-based approaches become analytically intractable or computationally prohibitive for these generalized asymmetric loss functions, our method provides explicit, manageable analytical forms. This unified framework effectively fills a gap in the literature, offering a systematic solution for smoothing a broad class of asymmetric non-smooth objective functions.

Extensive numerical experiments corroborate with theoretical analysis showing that the GMQ-based method significantly enhances computational efficiency through fast gradient-based optimization while maintaining high statistical accuracy.  Future research directions include extending this smoothing technique to nonlinear models, exploring its utility in smoothing activation functions (e.g., ReLU) in neural networks, and applying it to broader function approximation in high-dimensional statistics and machine learning.

\bibliographystyle{chicago-italics}
\bibliography{main}

@article{Hardy1971,
  title={Multiquadric equations of topography and other irregular surfaces},
  author={Rolland Lee Hardy},
  journal={Journal of Geophysical Research},
  year={1971},
  volume={76},
  pages={1905-1915},
  url={https://api.semanticscholar.org/CorpusID:129508657}
}

@article{He2020SmoothedQR,
  title={Smoothed Quantile Regression with Large-Scale Inference.},
  author={Xuming He and Xiaoou Pan and Kean Ming Tan and Wen-Xin Zhou},
  journal={Journal of econometrics},
  year={2020},
  volume={232 2},
  pages={367-388},
  url={https://api.semanticscholar.org/CorpusID:221835093}
}

@article{Koenker1978RQ,
 ISSN = {00129682, 14680262},
 URL = {http://www.jstor.org/stable/1913643},
 author = {Roger Koenker and Gilbert Bassett},
 journal = {Econometrica},
 number = {1},
 pages = {33--50},
 publisher = {[Wiley, Econometric Society]},
 title = {Regression Quantiles},
 urldate = {2023-11-02},
 volume = {46},
 year = {1978}
}

@article{Horowitz1996BootstrapMF,
  title={Bootstrap Methods for Median Regression Models},
  author={Joel L. Horowitz},
  journal={Econometrica},
  year={1996},
  volume={66},
  pages={1327-1351},
  url={https://api.semanticscholar.org/CorpusID:122629645}
}

@article{Barzilai1988TwoPointSS,
  title={Two-Point Step Size Gradient Methods},
  author={Jonathan Barzilai and Jonathan Michael Borwein},
  journal={Ima Journal of Numerical Analysis},
  year={1988},
  volume={8},
  pages={141-148},
  url={https://api.semanticscholar.org/CorpusID:123014714}
}

@article{Newey1987AsymmetricLS,
  title={Asymmetric Least Squares Estimation and Testing},
  author={Whitney Newey and James L. Powell},
  journal={Econometrica},
  year={1987},
  volume={55},
  pages={819-847},
  url={https://api.semanticscholar.org/CorpusID:40384446}
}

@article{Lin2022TheKP,
  title={The kth Power Expectile Estimation and Testing},
  author={Fuming Lin and Yingying Jiang and Yong Zhou},
  journal={Communications in Mathematics and Statistics},
  year={2022},
  volume={12},
  pages={573-615},
  url={https://doi.org/10.1007/s40304-022-00302-w}
}

@article{Fernandes2021SmoothQR,
    author = {Marcelo Fernandes, Emmanuel Guerre and Eduardo Horta},
    title = {Smoothing Quantile Regressions},
    journal = {Journal of Business  Economic Statistics},
    volume = {39},
    number = {1},
    pages = {338-357},
    year = {2021},
    publisher = {Taylor  Francis},
    doi = {10.1080/07350015.2019.1660177},
    URL = {https://doi.org/10.1080/07350015.2019.1660177}
}

@article{Galvao2016SmoothQRPanel,
    author = {Antonio F. Galvao, Kengo Kato},
    title = {Smoothed quantile regression for panel data},
    journal = {Journal of Econometrics},
    volume = {193},
    number = {1},
    pages = {92-112},
    year = {2016},
    doi = {10.1016/j.jeconom.2016.01.008},
    URL = {https://doi.org/10.1016/j.jeconom.2016.01.008}
}

@article{WANG2024ConvoSmoothForSVM,
    author = {Kangning Wang and Junning Yang and Kemal Polat and Adi Alhudhaif and Xiaofei Sun},
    title = {Convolution smoothing and non-convex regularization for support vector machine in high dimensions},
    journal = {Applied Soft Computing},
    volume = {155},
    pages = {111433},
    year = {2024},
    doi = {https://doi.org/10.1016/j.asoc.2024.111433},
    url = {https://www.sciencedirect.com/science/article/pii/S1568494624002072},
}

@article{Zhou02042024,
author = {Le Zhou and Boxiang Wang and Hui Zou},
title = {Sparse Convoluted Rank Regression in High Dimensions},
journal = {Journal of the American Statistical Association},
volume = {119},
number = {546},
pages = {1500--1512},
year = {2024},
publisher = {ASA Website},
doi = {10.1080/01621459.2023.2202433},
URL = {https://doi.org/10.1080/01621459.2023.2202433},
}

@article{Tan2021Convo,
    author = {Tan, Kean Ming and Wang, Lan and Zhou, Wen-Xin},
    title = {High-Dimensional Quantile Regression: Convolution Smoothing and Concave Regularization},
    journal = {Journal of the Royal Statistical Society Series B: Statistical Methodology},
    volume = {84},
    number = {1},
    pages = {205-233},
    year = {2021},
    month = {12},
    issn = {1369-7412},
    doi = {10.1111/rssb.12485},
    url = {https://doi.org/10.1111/rssb.12485},
}

@article{Liu2024,
title = {A Unified Inference for Predictive Quantile Regression},
author = {Liu, Xiaohui and Long, Wei and Peng, Liang and Yang, Bingduo},
year = {2024},
journal = {Journal of the American Statistical Association},
volume = {119},
number = {546},
pages = {1526-1540},
url = {https://EconPapers.repec.org/RePEc:taf:jnlasa:v:119:y:2024:i:546:p:1526-1540}
}

@article{He2025108128,
title = {Extremal local linear quantile regression for nonlinear dependent processes},
journal = {Computational Statistics  Data Analysis},
volume = {206},
pages = {108128},
year = {2025},
issn = {0167-9473},
doi = {https://doi.org/10.1016/j.csda.2025.108128},
url = {https://www.sciencedirect.com/science/article/pii/S0167947325000040},
author = {Fengyang He and Huixia Judy Wang}
}

@article{Yao2025,
  title={Graph-constrained quantile regression: Unifying structured regularization and robust modeling for enhanced accuracy and interpretability},
  author={Yao Dong and He Jiang and Sheng Panand Jianzhou Wang},
  journal={Information Sciences},
  year={2025},
  volume={720 },
  pages={Article 122530},
  doi={10.1016/j.ins.2025.122530}
}

@article{Takeuchi2006,
  author  = {Ichiro Takeuchi and Quoc V. Le and Timothy D. Sears and Alexander J. Smola},
  title   = {Nonparametric Quantile Estimation},
  journal = {Journal of Machine Learning Research},
  year    = {2006},
  volume  = {7},
  number  = {45},
  pages   = {1231--1264},
  url     = {http://jmlr.org/papers/v7/takeuchi06a.html}
}

@article{Zou2008Composite,
  title={Composite quantile regression and the oracle Model Selection Theory},
  author={ Zou, Hui  and  Yuan, Ming },
  journal={Annals of Stats},
  volume={36},
  number={3},
  pages={1108-1126},
  year={2008},
}

@article{MA2009925,
title = {Approximation to the k-th derivatives by multiquadric quasi-interpolation method},
journal = {Journal of Computational and Applied Mathematics},
volume = {231},
number = {2},
pages = {925-932},
year = {2009},
issn = {0377-0427},
doi = {https://doi.org/10.1016/j.cam.2009.05.017},
url = {https://www.sciencedirect.com/science/article/pii/S0377042709003227},
author = {Limin Ma and Zongmin Wu}
}

@article{Ma2010Stability,
  title={Stability of multiquadric quasi-interpolation to approximate high order derivatives},
  author={Limin Ma and Zongmin Wu},
  journal={SCIENCE CHINA Mathematics},
  year={2010},
  volume = {53},
  pages = {985-992}
}

@article{WuMa2011,
  title={Generator,multiquadric generator,quasi-interpolation and multiquadric quasi-interpolation},
  author={ Zong-min WU and Li-min MA },
  journal={ Applied Mathematics-A Journal of Chinese Universities},
  year={2011},
  volume = {26},
  pages = {390-400}
}

\appendix
\section{Proof of some lemmas and theorems}
\textbf{Proof of lemma \ref{theoreticalerror}}
\begin{proof}
Note that 
\begin{equation*}
\int|\rho_{\tau}(t)-\rho_{\tau,c}(t)|\mathrm{d}F(t;\beta(\tau))\leq \frac{1}{2}\int_{|t|\leq c}(\sqrt{c^2+t^2}-|t|)\mathrm{d}F(t;\beta(\tau))+\frac{1}{2}\int_{|t|\geq c}(\sqrt{c^2+t^2}-|t|)\mathrm{d}F(t;\beta(\tau)).
\end{equation*}
This together with Inequality \eqref{errorofsmooth} and the boundedness of  $f(t;\beta(\tau))=F'(t;\beta(\tau))$  leads to

\begin{equation*}
\begin{split}
\int|\rho_{\tau}(t)-\rho_{\tau,c}(t)|\mathrm{d}F(t;\beta(\tau))&\leq \frac{c}{2}C_1\int_{|t|\leq c}f(t;\beta(\tau))\mathrm{d}t+\frac{c^2}{2}C_2\left(\int_{c\leq |t|\leq 1}\frac{1}{|t|}f(t;\beta(\tau))\mathrm{d}t+\int_{1\leq |t|}f(t;\beta(\tau))\mathrm{d}t\right)\\
&\leq c^2(C_1+C_2|\ln c|)\|f\|_{\infty}+c^2C_2\\
&=\mathcal{O}(c^2|\ln c|).
\end{split}
\end{equation*}
Consequently, we have proven the lemma \ref{theoreticalerror}.
\end{proof}

\noindent\textbf{Proof of Lemma \ref{lemma:stochastic_order}}
\begin{proof}
Since $\hat{\beta}_c(\tau)$ is the unique minimizer of the smooth empirical objective function $R_{n,c}(\cdot)$, it satisfies the first-order condition $\nabla R_{n,c}(\hat{\beta}_c(\tau)) = \mathbf{0}$. Applying the first-order Taylor expansion with the integral remainder to $\nabla R_{n,c}$ around $\beta_c(\tau)$, we obtain the exact representation:
\begin{equation}\label{eq:expansion}
    \mathbf{0} = \nabla R_{n,c}(\hat{\beta}_c(\tau)) = \nabla R_{n,c}(\beta_c(\tau)) + H_n \big(\hat{\beta}_c(\tau) - \beta_c(\tau)\big),
\end{equation}
where $H_n := \int_0^1 \nabla^2 R_{n,c} \left( \beta_c(\tau) + t[\hat{\beta}_c(\tau) - \beta_c(\tau)] \right) \mathrm{d}t$ is the integrated sample Hessian matrix.

Due to the convexity of the GMQ loss function, the sample Hessian $\nabla^2 R_{n,c}(b)$ is positive definite for any $b$. Consequently, $H_n$ is invertible. Rearranging \eqref{eq:expansion} yields:
\begin{equation*}
    \hat{\beta}_c(\tau) - \beta_c(\tau) = - H_n^{-1} \nabla R_{n,c}(\beta_c(\tau)).
\end{equation*}
Given the consistency of $\hat{\beta}_c(\tau)$, for sufficiently large $n$, $\|H_n^{-1}\|$ is bounded by a positive constant $C_H$ with probability approaching one. Taking the spectral norm on both sides, we have:
\begin{equation}\label{eq:norm_inequality}
    \left\|\hat{\beta}_c(\tau) - \beta_c(\tau)\right\| \leq \|H_n^{-1}\| \left\|\nabla R_{n,c}(\beta_c(\tau))\right\| \leq C_H \left\|\nabla R_{n,c}(\beta_c(\tau))\right\|.
\end{equation}
Furthermore, following the rigorous framework established in Fernandes \cite{Fernandes2021SmoothQR}, the score function satisfies the exponential tail bound:
\begin{equation*}
    \mathbb{P}\left( \left \|\sqrt{n}\nabla R_{n,c}(\beta_c(\tau))\right \|\ge C_1(1+r) \right) \leq C_0 \exp(-r^2).
\end{equation*}
The algebraic bound in \eqref{eq:norm_inequality} dictates that if the estimation error $\left\|\hat{\beta}_c(\tau) - \beta_c(\tau)\right\|$ exceeds a threshold $\delta$, the scaled score function $C_H \left\|\nabla R_{n,c}(\beta_c(\tau))\right\|$ must necessarily exceed the same threshold. Consequently, for $\delta = \frac{C_H C_1(1+r)}{\sqrt{n}}$, we have:
\begin{align*}
    \mathbb{P}\left( \left\|\hat{\beta}_c(\tau) - \beta_c(\tau)\right\| \ge \delta \right) 
    &\le \mathbb{P}\left( C_H \left\|\nabla R_{n,c}(\beta_c(\tau))\right\| \ge \delta \right) \\
    &= \mathbb{P}\left( \left\|\sqrt{n}\nabla R_{n,c}(\beta_c(\tau))\right\| \ge C_1(1+r) \right) \\
    &\le C_0 \exp(-r^2).
\end{align*}
This tail bound immediately implies the root-$n$ consistency:
\begin{equation*}
    \left\|\hat{\beta}_c(\tau) - \beta_c(\tau)\right\| = O_p(n^{-1/2}).
\end{equation*}
\end{proof}

\end{document}